\documentstyle[psfig]{mn}

\def\etal{et~al.}
\def\spose#1{\hbox to 0pt{#1\hss}}
\def\lta{\mathrel{\spose{\lower 3pt\hbox{$\mathchar"218$}}
     \raise 2.0pt\hbox{$\mathchar"13C$}}}
\def\gta{\mathrel{\spose{\lower 3pt\hbox{$\mathchar"218$}}
     \raise 2.0pt\hbox{$\mathchar"13E$}}}
\def\Ha{H$\alpha$}
\def\Hb{H$\beta$}
\def\Hc{H$\gamma$}
\def\Hd{H$\delta$}

\def\iraf{{\sc iraf}}
\def\noao{{\sc noao}}

\def\kms{\,km\,s\,$^{-1}$}
\def\Ho50{$H_0 = 50$km\,s$^{-1}$\,Mpc$^{-1}$}
\def\ecs{\,erg\,s$^{-1}$\,cm$^{-2}$}
\def\ecsa{\,erg\,s$^{-1}$\,cm$^{-2}$\,\AA$^{-1}$}

\title[Spectroscopy of distant 3CR radio galaxies]{Deep spectroscopy of
distant 3CR radio galaxies: the data}

\author[P.~N.~Best \etal]{P.~N.~Best,$^1$\thanks{Email: pbest@strw.leidenuniv.nl} H.~J.~A.~R\"ottgering$^1$ and
M.~S.~Longair$^2$\\ 
$^1$ Sterrewacht Leiden, Postbus 9513, 2300 RA Leiden, the Netherlands\\ 
$^2$ Cavendish Astrophysics, Madingley Road, Cambridge, CB3 0HE, UK}

\pagerange{\pageref{firstpage}--\pageref{lastpage}}
\pubyear{1999}
\begin{document}
\label{firstpage}

\maketitle

\begin{abstract}
\noindent Deep long--slit spectroscopic data are presented for a sample of
14 3CR radio galaxies at redshift $z \sim 1$, previously studied in detail
using the Hubble Space Telescope, the Very Large Array, and UKIRT.
Analysis of the [OII]~3727 emission line structures at $\sim 5$\AA\
spectral resolution is carried out to derive the kinematic properties of
the emission line gas. In line with previous, lower resolution, studies, a
wide variety of kinematics are seen, from gas consistent with a mean
rotational motion through to complex structures with velocity dispersions
exceeding 1000\kms. The data confirm the presence of a high velocity gas
component in 3C265 and detached emission line systems in 3C356 and 3C441,
and show for the first time that the emission line gas in the central
regions of 3C324 is composed of two kinematically distinct
components. Emission line fluxes and the colour of the continuum emission
are determined down to unprecedently low observed wavelengths, $\lambda <
3500$\AA, sufficiently short that any contribution of an evolved stellar
population is negligible.  An accompanying paper investigates the
variation in the emission line ratios and velocity structures within the
sample, and draws conclusions as to the origin of the ionisation and
kinematics of these galaxies.
\end{abstract}

\begin{keywords}
Galaxies: active --- Galaxies: interstellar medium --- Radio continuum:
galaxies --- Galaxies: Individual: 3C324, 3C265
\end{keywords}

\section{Introduction}

The emission line properties of powerful distant ($z \gta 0.5$) radio
galaxies are striking. Their emission line luminosities are large, with
the rest--frame equivalent width of the [OII]~3727 line frequently
exceeding 100\AA\ (e.g. Spinrad 1982)\nocite{spi82}. Indeed, the strong
correlation between emission line luminosity and radio power (e.g. Rawlings \&
Saunders 1991)\nocite{raw91b} was the key factor in enabling spectroscopic
completeness to be achieved for a large sample of powerful radio galaxies
(the revised 3CR catalogue; Laing \etal\ 1983)\nocite{lai83}. The line
emission of the distant 3CR radio galaxies is also seen to be spatially
extended over regions that can be as large as 100\,kpc and is frequently
elongated along the direction of the radio axis (e.g. McCarthy 1988,
McCarthy \etal\ 1995).\nocite{mcc88,mcc95}

The source of ionisation of this gas has been a long standing question.
Robinson et~al. \shortcite{rob87} found that optical emission line spectra
of most low redshift ($z \lta 0.1$) radio galaxies are well explained
using photoionisation models, and a similar result was found for a
composite spectrum of radio galaxies with redshifts $0.1 < z < 3$
\cite{mcc93}.  Photoionisation models are also supported by
orientation--based unification schemes of radio galaxies and radio--loud
quasars (e.g. Barthel 1989)\nocite{bar89}, in which all radio galaxies
host an obscured quasar nucleus: the flux of ionising photons required to
produce the observed luminosities of the emission line regions can be
shown to be comparable to that produced by radio--loud quasars at the same
redshift (e.g. see McCarthy 1993)\nocite{mcc93}. On the other hand,
detailed studies of individual sources (e.g. 3C277.3; van Breugel et~al
1985; 3C171; Clark et~al. 1998)\nocite{bre85b,cla98} have revealed
features such as enhanced nebular line emission, high velocity gas
components, and large velocity dispersions coincident with the radio
hotspots or with bends in the radio jets, indicating that the morphology
and kinematics of the gas in some sources are dominated by shocks
associated with the radio source. The ionisation state of the gas in these
regions is also consistent with that expected from shock ionisation
(e.g. Villar--Mart{\'i}n \etal\ 1999\nocite{vil99a}). Bicknell \etal\
\shortcite{bic97} considered the energy input to the emission line regions
of Gigahertz-Peaked Spectrum (GPS) and Compact Steep Spectrum (CSS)
sources from the shocks associated with the advance of the radio jet and
cocoon, and showed that the energy supplied by the shocks to the
interstellar medium is sufficient to account for the observed line
emission. The relative importance of shocks and photoionisation in
producing the emission line properties of the general radio galaxy
population therefore remains an open question.

Another important issue is the varied kinematics seen in the emission line
regions. At low redshifts the emission line properties of the 3CR radio
galaxies have been intensively studied (e.g. Baum \etal\ 1992 and
references therein)\nocite{bau92}; a variety of kinematics are seen, from
galaxies consistent with simple rotation through to those classified as
`violent non-rotators' with large turbulent velocities.  At higher
redshifts, McCarthy \etal\ \shortcite{mcc95,mcc96a} have studied a large
sample of 3CR radio galaxies with low spectral and spatial resolution, and
find that the velocity full--width--half--maxima (FWHM) are significantly
higher than those at low redshifts (see also Baum \etal\
1998)\nocite{bau98b}, often exceeding 1000\kms, and large velocity shears
are seen. The exceptional nature of the kinematics has been reinforced by
more detailed studies of individual sources (e.g. Spinrad \& Djorgovski
1984, Tadhunter 1991, Meisenheimer \& Hippelein 1992, Hippelein \&
Meisenheimer 1992, Stockton \etal\ 1996, Neeser \etal\
1997)\nocite{spi84a,tad91,mei92,hip92,sto96a,nee97b}. The emission line
properties of these high redshift radio galaxies are evidently more
extreme than those at low redshift (and hence of lower radio power) in
more than just their luminosities.

The origin of the emission line gas itself is another unresolved
issue. Typically $10^8$ to $10^9 M_{\odot}$ of ionised gas are estimated
to be present around these objects (McCarthy 1993 and references
therein)\nocite{mcc93}, significantly more than found in quiescent low
redshift ellipticals. The gas may have an origin external to the radio
galaxy, being either associated with the remnants of a galaxy merger
\cite{hec86,bau89b}, or gas brought in by a massive cooling flow in a
surrounding intracluster medium; some support for the latter hypothesis is
given by the detection of extended X--ray emission around a number of
powerful distant radio galaxies (Crawford and Fabian 1996 and references
therein)\nocite{cra96b}, although the higher than primordial metallicity
of the gas (as indicated by the strong emission lines of, for example,
oxygen, neon, magnesium and sulphur) dictates that the gas must have have
been processed within stars at some point in its past. Alternatively, the
gas may be left over from the formation phase of these massive galaxies,
perhaps expelled from the galaxy either in a wind following an earlier
starburst phase or more recently by the shocks associated with the radio
source. If the gas has an origin external to the host galaxy, then it is
important to know what the connection is, if any, between the origin of
this gas and the onset of the radio source activity.

The properties of the continuum emission of powerful distant radio
galaxies are equally interesting. At near infrared wavelengths the
galaxies follow a tight K$-z$ Hubble relation \cite{lil84a} and their host
galaxies have colours and radial light profiles consistent with being
giant elliptical galaxies which formed at large redshifts
\cite{bes98d}. At optical wavelengths, however, powerful radio galaxies
beyond redshift $z \sim 0.6$ show a strong, but variable, excess of blue
emission, generally aligned along the radio axis \cite{mcc87,cha87}. Using
Hubble Space Telescope (HST) images of a sample of 3CR radio galaxies with
redshifts $z \sim 1$, we have shown that the nature of this alignment
differs greatly from galaxy to galaxy, in particular becoming weaker as
the linear size of the radio source increases \cite{bes96a,bes97c}. It is
clear that a number of different physical processes contribute to the
continuum alignment effect, but less clear which processes are the most
important (for reviews see e.g. McCarthy 1993, R{\"o}ttgering \& Miley
1996).\nocite{mcc93,rot96d}

To study the emission line gas properties of these galaxies, our
multi--waveband imaging project on the redshift one 3CR galaxies has been
expanded to include deep spectroscopic observations, producing a combined
dataset of unparalleled quality.  In the current paper the basic results
of the spectroscopic program are presented. The layout is as follows. In
Section~\ref{observ}, details concerning the sample selection, the
observations and the data reduction are presented.  Section~\ref{results}
contains the direct results of these observations, in the form of
extracted one--dimensional spectra, two--dimensional studies of the
[OII]~3727 emission line structures, tables of spectral properties, and a
brief description of the individual sources. The results are summarised in
Section~\ref{conc}. An accompanying paper (Best et~al 1999; hereafter
Paper 2)\nocite{bes99c} investigates the emission line ratios and velocity
structures of the sample as a whole, and the consequences of these for the
origin of the ionisation and kinematics of these galaxies. A later paper
will address the nature of the continuum emission.

Throughout the paper, values of the cosmological parameters of $\Omega =
1$ and $H_0 = 50$\kms\,Mpc$^{-1}$ are assumed. For this cosmology, 1
arcsec corresponds to 8.5\,kpc at redshift $z=1$.

\section{Observational Details}
\label{observ}

\subsection{Sample selection and observational set-up}
\label{sample}

The galaxies were drawn from the essentially complete sample of 28 3CR
radio galaxies with redshifts $0.6 < z < 1.8$ which we have
intensively studied using the HST, the VLA and UKIRT (e.g. Best \etal\
1997).\nocite{bes97c} From this sample, spectroscopic studies were
restricted initially to those 18 galaxies with redshifts $0.7 < z <
1.25$, the upper redshift cut--off corresponding to that at which the
4000\AA\ break is redshifted beyond an observed wavelength of 9000\AA\
and the lower redshift cut-off being set by telescope time
limitations. Of these 18 galaxies, 3C41 ($z=0.795$), 3C65 ($z=1.176$),
3C267 ($z=1.144$) and 3C277.2 ($z=0.766$) were subsequently excluded. In
the cases of 3C65 and 3C267, their exclusion was due to partially
cloudy conditions during one observing night resulting in a poor data
quality for these two observations. The omission of 3C41 and 3C277.2
was due to constraints of telescope time at the relevant right
ascensions: the decision not to observe these particular two galaxies
was based solely upon them having the lowest redshifts in the sample
at those right ascensions. The exclusion of these four galaxies should
not introduce any significant selection effects.

The remaining 14 galaxies were observed on July 7-8 1997 and February
23-24 1998, using the duel--beam ISIS spectrograph on the William Herschel
Telescope (WHT). The 5400\AA\ dichroic was selected since it provided the
highest throughput at short wavelengths. The R158B grating was used in
combination with a Loral CCD in the blue arm of the spectrograph. This low
spectral resolution ($\sim 12$\AA) grating provided the largest wavelength
coverage (3700\AA) and maximized the signal--to--noise at short
wavelengths by decreasing the importance of the CCD read-out noise, which
even still was the dominant source of noise at wavelengths below about
3400\AA. This set--up enabled accurate measurement of many emission line
strengths and a determination of the slope of the continuum emission at
sufficiently short wavelengths that any contribution of the evolved
stellar population will be negligible; at such wavelengths the spectral
characteristics of the aligned emission alone are measured and can be used
to pin down the physical processes contributing to the alignment effect.
During the July run, the wavelength range sampled by the CCD was set to
span from below the minimum useful wavelength ($\sim 3250$\AA) to longward
of the dichroic. During the February run, the (different) Loral CCD had a
charge trap in the dispersion direction at about pixel 1000, reducing the
credibility of data at longer wavelengths; the wavelength range was tuned
to sample from 3275\AA\ up to the charge trap at about 5100\AA.

In the red arm of the spectrograph the R316R grating was used in
combination with TEK CCD, providing a spatial scale of 0.36 arcsec per
pixel, a dispersion of 1.49\AA\ per pixel and a spectral resolution of
about 5\AA. The wavelength range of about 1500\AA\ was centred on the
wavelength given in Table~\ref{obstab} for each galaxy, tuned to cover as
much as possible of the range from approximately 3550\AA\ to 4300\AA\ in
the rest--frame of the galaxy whilst remaining below a maximum observed
wavelength of 9000\AA. This higher spectral resolution set-up in the red
arm allows a much more detailed investigation of the velocity structures
of the emission line gas as seen in the very luminous [OII]~3727 emission
line, but still provides sufficient wavelength coverage to include the
4000\AA\ break, the Balmer continuum break at 3645\AA\, and a number of
Balmer emission lines.

\begin{table*}
\caption{\label{obstab} Details of the ISIS observations.}
\begin{center}
\begin{tabular}{lccccccc}
Source & Observ. &Slit width& Exp. Time  & Exp. Time   & Red Arm   & Slit PA & Notes \\
       &   Date  &  [$''$]  & Red Arm [s]& Blue Arm [s]&Cen. $\lambda$ [\AA]&[deg.]&    \\
3C22   & 07/07/97&  1.50    &  5400      &  5400       &  7645     &   103   & [1]   \\
3C217  & 23/02/98&  1.54    &  7250      &  7200       &  7625     &   90    & [2,3] \\
3C226  & 23/02/98&  1.54    &  7200      &  7200       &  7260     &   145   & [1]   \\
3C247  & 24/02/98&  1.70    &  6600      &  6660       &  7100     &   70    & [1]   \\
3C252  & 24/02/98&  1.75    &  7140      &  7200       &  8235     &   105   & [1]   \\
3C265  & 23/02/98&  1.54    &  5500      &  5500       &  7260     &   136   & [4]   \\
3C280  & 07/07/97&  1.50    &  5400      &  5400       &  7860     &    90   & [1]   \\
3C289  & 08/07/97&  1.50    &  2700      &  2700       &  7760     &   109   & [1]   \\
3C289  & 23/02/98&  1.54    &  2200      &  2200       &  7820     &   109   & [1]   \\
3C324  & 07/07/98&  1.50    &  5400      &  5400       &  7860     &    90   & [1]   \\
3C340  & 24/02/98&  1.75    &  6900      &  7000       &  7105     &    85   & [1]   \\
3C352  & 08/07/98&  1.50    &  5400      &  5400       &  7170     &   161   & [1]   \\
3C356  & 07/07/98&  1.50    &  5400      &  5400       &  7860     &   147   & [5]   \\
3C368  & 08/07/98&  1.50    &  5400      &  5400       &  8235     &    10   & [1]   \\
3C441  & 08/07/98&  1.50    &  5400      &  5400       &  6830     &   150   & [1]   \\
\end{tabular}
\end{center}

\raggedright [1] Slit aligned along the radio axis. \\ \raggedright [2]
Slit aligned intermediately between the radio axis (104$^{\circ}$) and the
elongation of the central optical knots (75$^{\circ}$). \\ \raggedright
[3] Observations may be non--photometric (see text). \\ \raggedright [4]
Slit aligned along the extended UV emission, rather than the radio axis
which has a PA of 106$^{\circ}$. \\ \raggedright [5] Slit aligned to
include both of the two central galaxies; slightly offset from the true
radio PA of 160$^{\circ}$.

\end{table*}

\subsection{Observations and data reduction}
\label{reduc}

Long--slit spectra of the 14 galaxies were taken with total integration
times of between 1.5 and 2 hours per galaxy; the observations were split
between 3 or 4 separate exposures in the red arm to assist in the removal
of cosmic rays; the blue arm observations were split between only 2
exposures since shorter exposures would have had a more significant
read--noise contribution. The slit was orientated either along the radio
axis or along the axis of elongation of the optical--UV emission. Full
details of the observations are provided in Table~\ref{obstab}.

The seeing was typically 0.8 to 1 arcsec during the July run, and
between $1$ and $1.25$ arcsec during the February observations. The
first half of the February 23 night was partially cloudy, hampering
the observations of 3C65 and 3C267 as discussed above. The
observations of 3C217 {\it may} have suffered partial cloud
interference and be non--photometric, although the approximate
agreement between the 7500\AA\ flux density determined from the
spectrum and that extracted from the equivalent region of an HST image
of this galaxy, convolved to the same angular resolution, suggest that
this was not significant.  Conditions during the second half of that
night and the other three nights were photometric.
 
The data were reduced using standard packages within the \iraf\ \noao\
reduction software. The raw data frames were corrected for overscan bias,
and flat--fielded using observations of internal calibration lamps with
the same instrumental set-up as the object exposures: ie. in the red arm,
separate flat fields were constructed for each galaxy, since each was
observed with a different {\it observed} wavelength range; these flat
field observations were interspersed with the series of on--source
exposures to minimise fringing effects. The sky background was removed,
taking care not to include extended line emission in the sky bands. The
different exposures of each galaxy were then combined, removing the cosmic
ray events, and one dimensional spectra were extracted. The data were
wavelength calibrated using observations of CuNe and CuAr arc lamps, and
accurate flux calibration was achieved using observations of the
spectrophotometric standard stars GD190, EG79, G9937 and LDS749b, again
observed using exactly the same instrumental setup as each galaxy and
corrected for airmass extinction.

\begin{figure}
\centerline{
\psfig{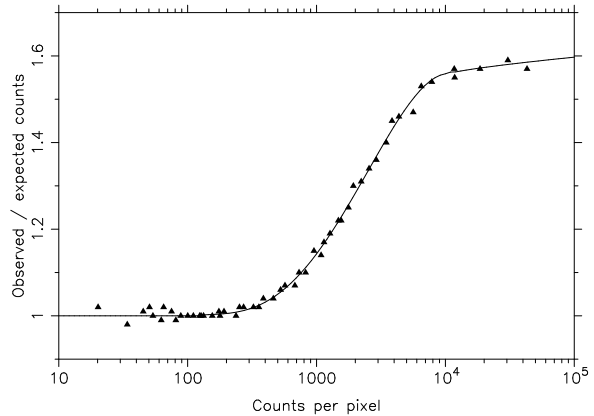}
}
\caption{\label{lorallin} The non-linearity of the Loral CCD during the
July run. The `expected' counts were determined by defining (arbitrarily)
for a given CCD region the frame in the exposure time sequence of flat
field frames (see text) that gave just over 100 counts per pixel to be
`correct', and scaling the expected counts for other frames by their
exposure time; hence the string of 1.00's from 100 to 150 counts.  The solid
line represents a polynomial fit to the data points.}
\end{figure}

\subsection{The non-linearity of the Loral CCD}
\label{loral}

The far greater efficiency of the Loral CCD at wavelengths below 4000\AA\
than any other CCD available at the time of these observations ($\approx
75$\% as compared with $\approx 35$\% quantum efficiency at 3500\AA)
offered an unrivalled opportunity for study at these wavelengths. However,
the Loral CCD used during the July run had a slightly non--linear response
curve, giving rise to a minor problem concerning flux calibration of the
blue arm data from this run. In order to assess the extent of this
problem, a sequence of flat field observations of the internal calibration
lamps were taken, with exposures of 0, 1, 2, 3, 5, 7, 10, 15, 20, 30, 60,
120, 300, 120, 60, 30, 20, 15, 10, 7, 5, 3, 2, 1, 0 seconds, the average
of the frames for each exposure time on the increasing and decreasing
exposure time branches being taken to account for any systematic time
variation in the intensity of the calibration lamp. A small (relatively
uniform intensity) region of the CCD was selected and, after subtraction
of the 0 second bias frame, the mean counts per pixel in that region was
measured for each different exposure time frame. The frame providing on
average just over 100 counts per pixel was arbitrarily declared to be
`correct'; the `expected' count level for this CCD region in the other frames
was then calculated by scaling by the exposure time, and compared to the
observed counts. This process was repeated for a large number of different
regions on the CCD, and also with a smaller sample of flat--field
observations taken the following night.

The results of this analysis are presented in Figure~\ref{lorallin}, which
shows that the Loral CCD is non--linear at the $\lta 10$\% level at count
levels below $\sim 800$ counts per pixel, but above that level the
non--linearity increases sharply. The scatter around a parameterised fit
to the non--linearity curve in the 300 to 500 counts range can be
explained by Poisson noise statistics alone, indicating that the
non--linearity was highly repeatable, varying neither with time nor with
position on the CCD, and thus allowing the small non--linearity at these
count levels to be accurately calibrated out.

The faintness of the radio galaxies being studied meant that the detected
counts per pixel in the on--source exposures, including both sky and
object counts, fell automatically within the range 50 to 800 counts per
pixel. Both the flat field and the standard star exposures in the blue arm
during the July run were built up by summing a series of images, each of
which was kept short to have maximum count levels below 1000 counts per
pixel. The parameterised curve shown in Figure~\ref{lorallin} was then
divided into the scientific exposures, after bias subtraction but before
flat fielding and other calibration. Such a correction was also applied to
the flat--field exposures and to the observations of standard stars. In
this way, any systematic offset introduced by the application of the
non-linearity curve to the object will be roughly cancelled by its
application to the calibrator, reducing any errors to $\lta 2$\%, far
below the other uncertainties related to the calibration procedure.

% Problems making Table 2 in landscape within MN style file => make it in
% landscape elsewhere and import it in as a figure!

\begin{figure*}
\centerline{
\psfig{file=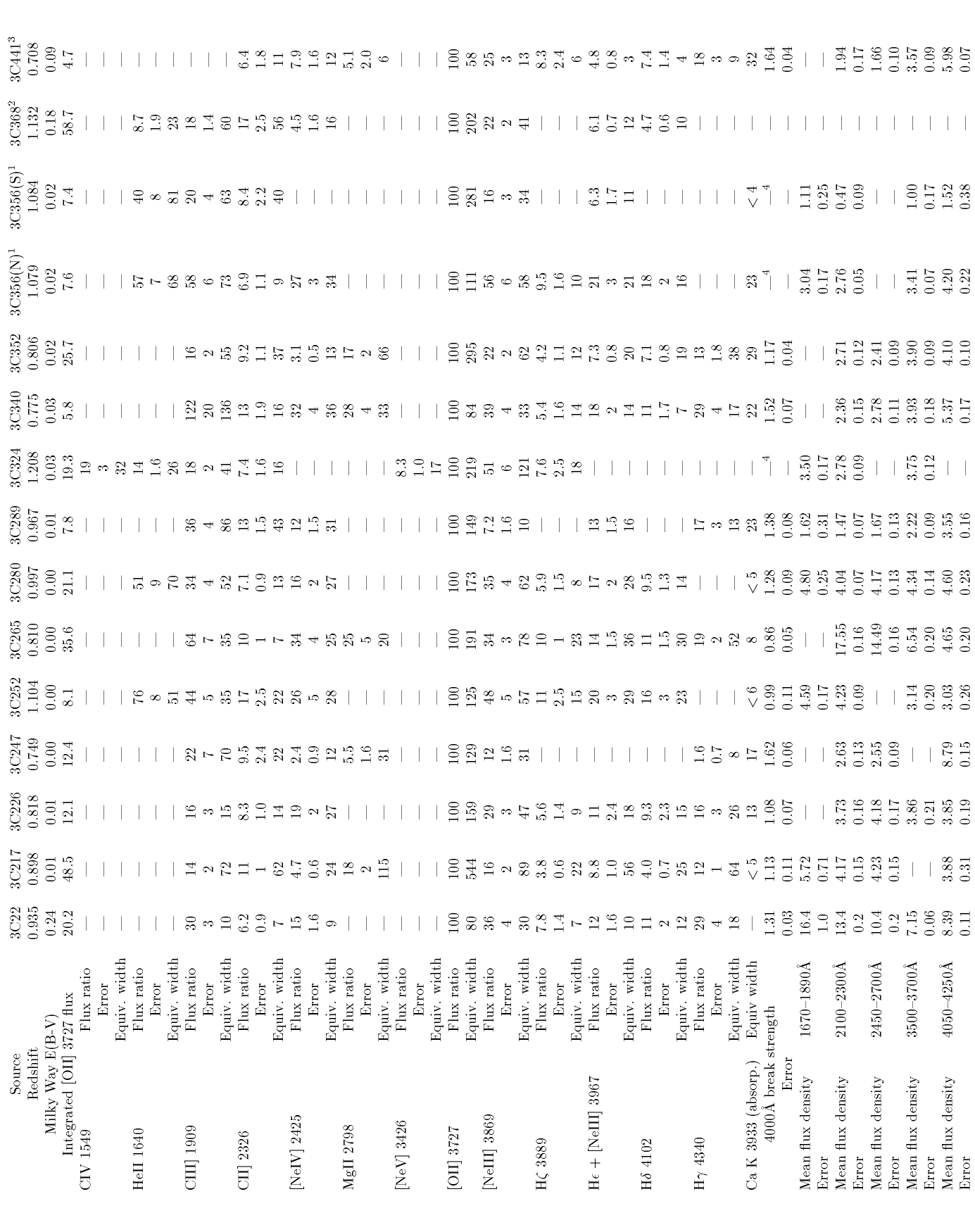,width=\textwidth,clip=}
}
\end{figure*}

\begin{table}
\caption{Spectroscopic properties of the radio galaxies. The [OII]~3727
integrated flux (measured within or scaled to a 1.5 arcsec slit width), in
units of $10^{-16}$\,\ecs, corresponds to the line flux along the entire
length of the slit calculated by integrating the [OII]~3727 intensities
shown in Figures 2-15(d). All of the flux ratios and flux densities quoted
in this table are corrected for galactic extinction using the E(B-V) for
the Milky Way from Burstein \& Heiles (1982) and the parameterised
galactic extinction law of Howarth (1983). All flux ratios are measured
relative to [OII]~3727 (value of 100) within the extracted 1--dimensional
spectrum. The error on the [OII]~3727 line flux is dominated by
calibration errors which are estimated to be $\lta 10$\%. Errors in the
other flux ratios take account of both the errors due to photon statistics
and a 10\% calibration uncertainty. Note that there might be a small $\lta
5-10$\% systematic offset between the blue and red arm measurements due to
different spatial extraction regions (the spatial extraction length is the
same in both arms, but might be centred slightly differently. Also, since
the slit position angles were different from the parallactic angle,
differential refraction may result in the galaxy not being completely
centred in the slit at both the red and blue extremes. This latter effect
should be minimal, however, as it was ensured that all observations were
taken either at low airmass or with the slit position angle close to the
parallactic angle). Equivalent widths are measured in the rest--frame of
the galaxy. Mean flux densities for given line--free wavelength ranges (or
as much of the range as possible provided at least 100\AA\ are covered in
the spectrum) are measured from the extracted one--dimensional spectrum;
the (weak) H$\delta$ emission line was subtracted before calculating the
mean continuum level for the wavelength range 4050-4250\AA. Values are in
units of $10^{-18}$\ecsa; the uncertainty given is the error on the mean
value in that wavelength region. Notes: [1] values are quoted separately
for the northern and southern host galaxy candidates for 3C356; [2] for
3C368, due to the presence of a galactic M-star coincident with the galaxy
(Hammer \etal\ 1991), the equivalent widths are strictly lower limits and
continuum flux densities have not been measured; [3] for 3C441 the values
quoted do not include the companion galaxy; [4] the red arm spectra do not
reach sufficiently high rest--frame wavelengths.}
\end{table}
\nocite{bur82a,how83,ham91}

\section{Results}
\label{results}

The resulting one dimensional spectra, extracted from the central 4.3
arcsec ($\sim 35$\,kpc) region along the slit direction in each of the
blue and red arms, are shown in Figures 2 to 15 (a \& b). In Table~2 are
tabulated the fluxes of the various emission lines relative to [OII]~3727
and their equivalent widths, together with the mean flux density of the
continuum in various wavelength regions. These flux ratios and flux
densities (although not the plotted one-dimensional spectra, to allow
comparison with previously published data) have been corrected for
galactic extinction using the Milky Way HI column density data of Burstein
and Heiles \shortcite{bur82a}, quoted in Table~2, and the parameterised
galactic extinction law of Howarth \shortcite{how83}. These extinction
corrections are $\lta 10$\% for most sources, but exceed a factor of 2 at
the shortest wavelengths for the low galactic latitude source 3C22.

The emission line flux ratios and continuum flux densities are tabulated
only for the single extracted spectrum. Even these very deep spectra do
not have high enough signal--to--noise in the blue continuum of most of
the galaxies to investigate in detail variations in the continuum colour
along the spatial direction of the slit. Variations in the intensity,
velocity and FWHM of the emission lines along the spatial direction of the
slit are readily apparent, and are considered below in the study of the
[OII]~3727 emission line.

\begin{figure*}
\centerline{
\psfig{file=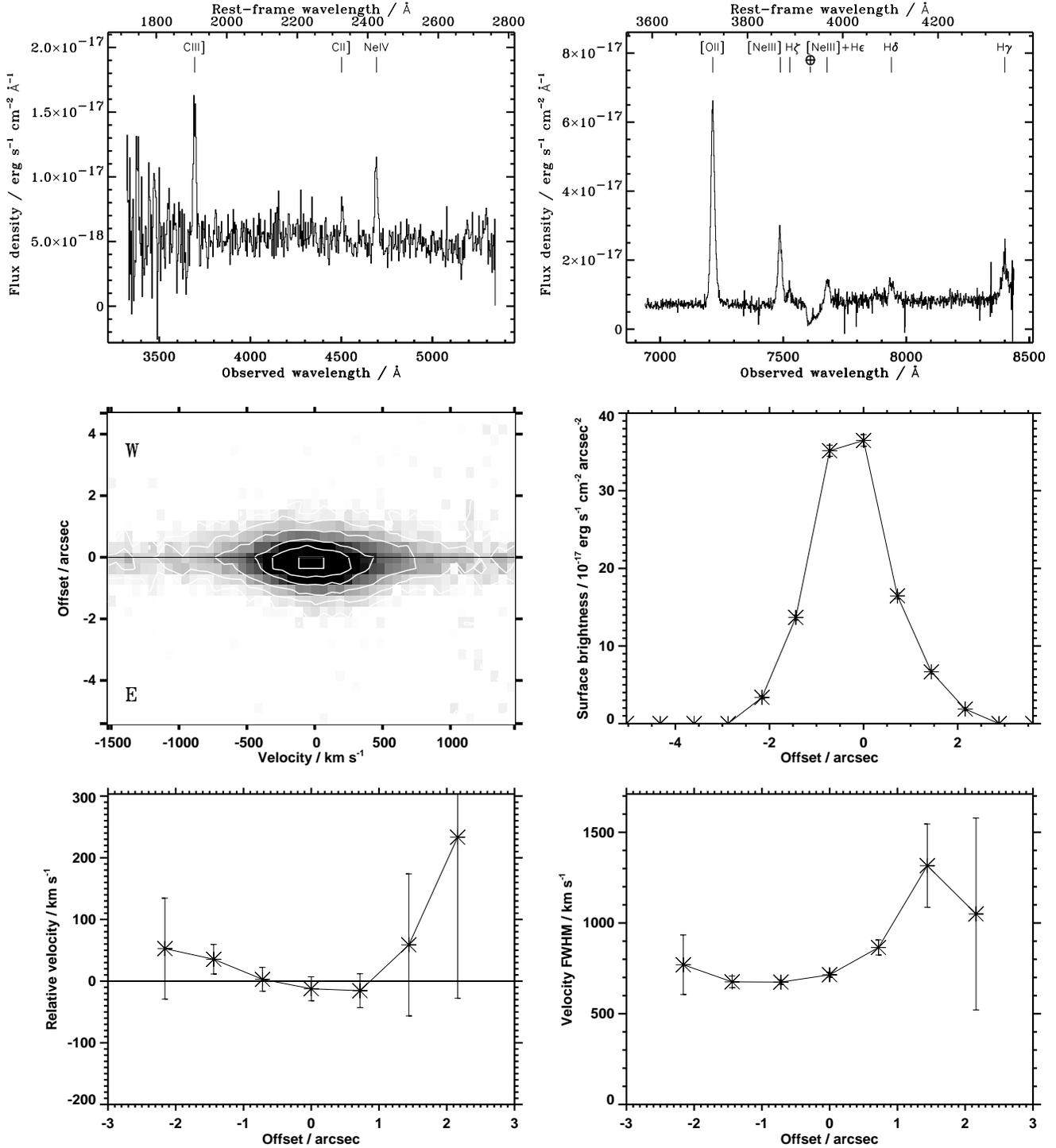,width=\textwidth,clip=} 
}
\caption{The spectroscopic data for {\bf 3C22}. (a -- upper left) The
extracted 1-dimensional blue arm spectrum. In this and in (b) the
emission lines are labelled and sky features are indicated by an open
circle with a cross. (b -- upper right) The 1-dimensional
spectrum extracted from the red arm. (c -- middle left) The
2-dimensional [OII]~3727 emission line structure. Offset zero
corresponds to the continuum centroid, and the sky directions of
`positive' and `negative' slit offsets are indicated on the plot. (d
-- middle right) The surface brightness of the [OII]~3727 emission as
a function of position along the slit. Where a second Gaussian
component is fitted (see text; appropriate for later figures) this is
plotted using open diamonds. (e -- lower left) The velocity at the
Gaussian peak of the [OII]~3727 emission as a function of position
along the slit. (f -- lower right) The variation of the FWHM of the
fitted Gaussian profile as a function of position along the slit.}
\end{figure*}

\begin{figure*}
\centerline{
\psfig{file=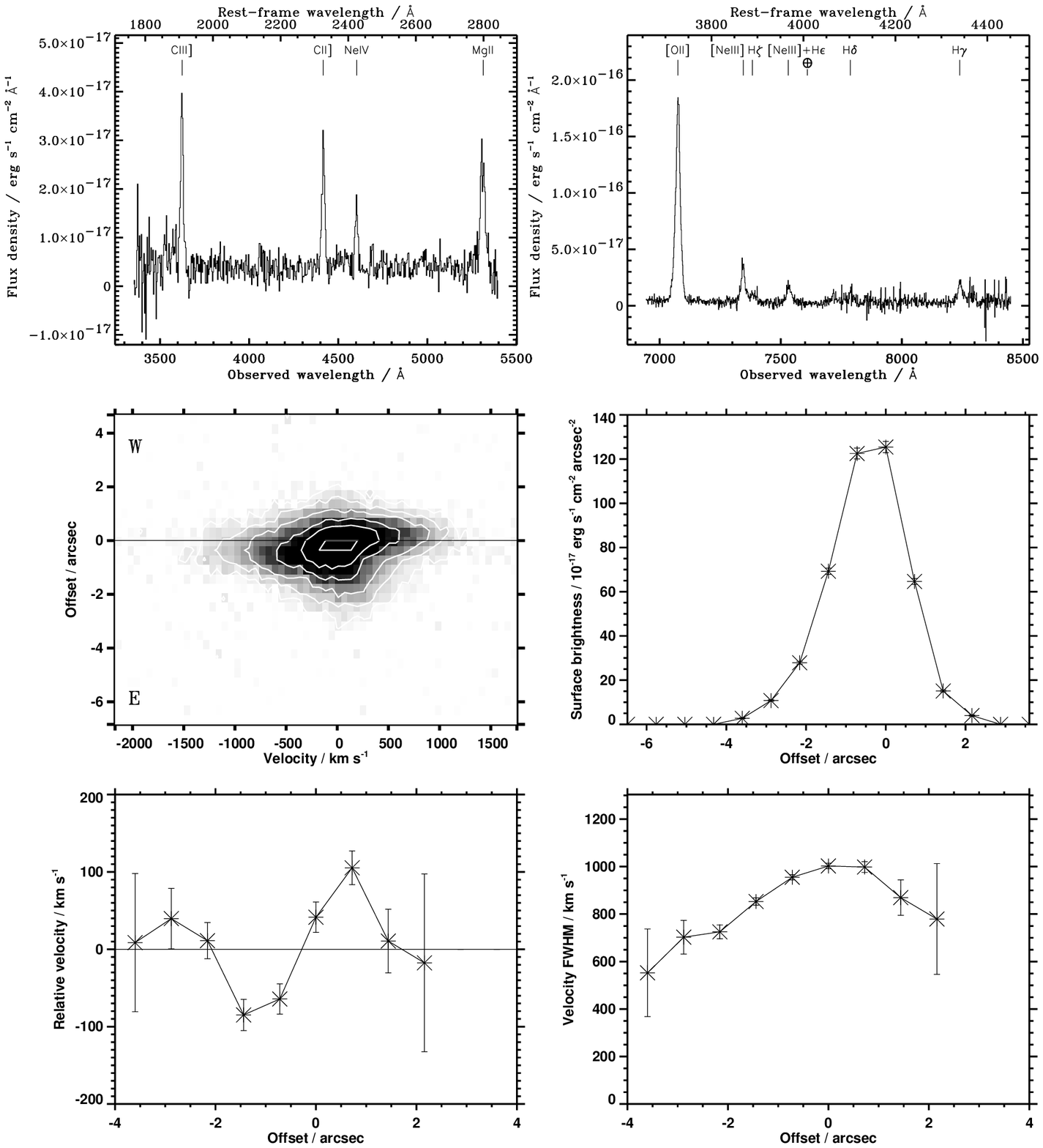,width=\textwidth,clip=}
}
\caption{The spectroscopic data for {\bf 3C217}. Details as in Figure~2.}
\end{figure*}

\begin{figure*}
\centerline{
\psfig{file=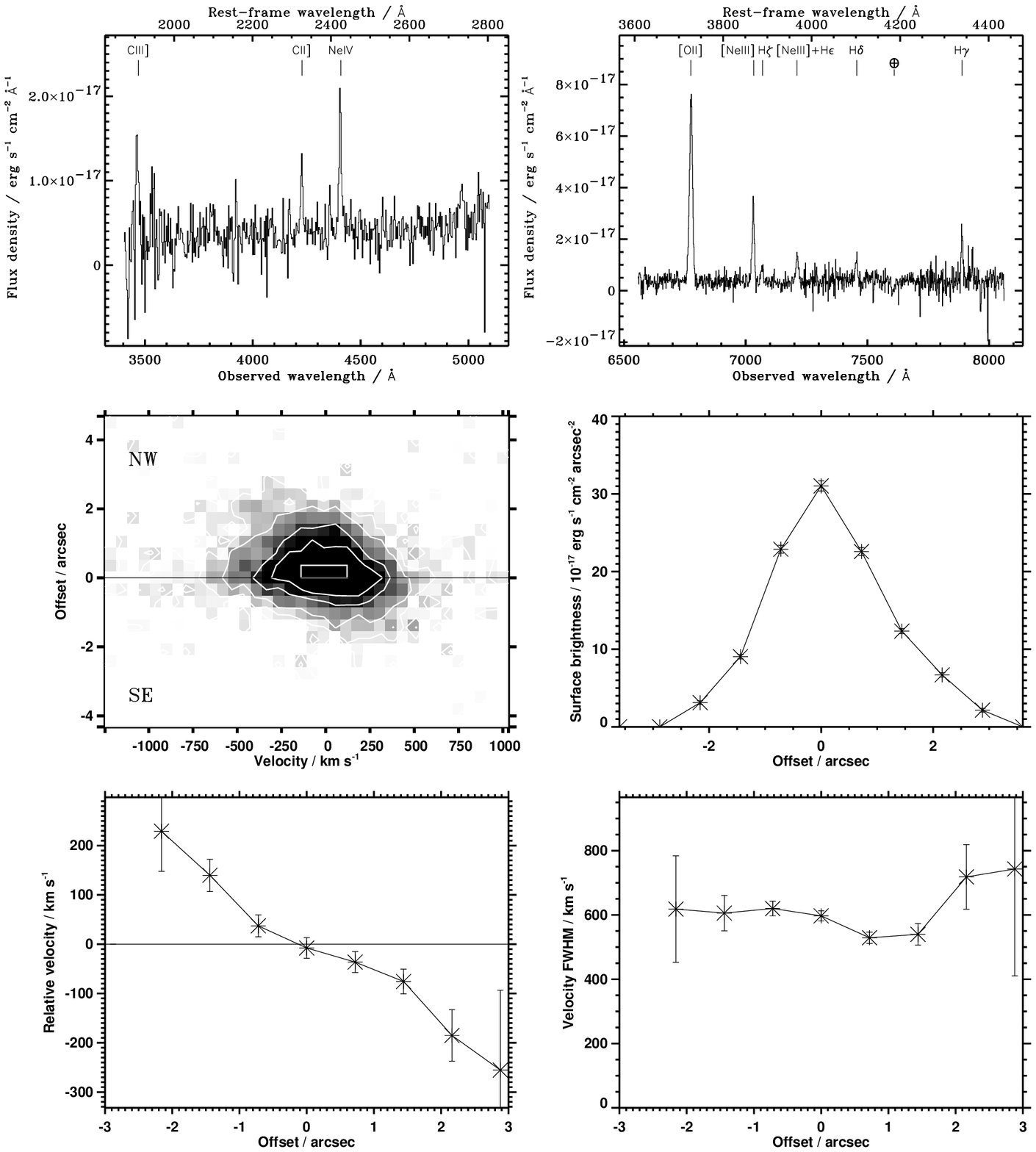,width=\textwidth,clip=}
}
\caption{The spectroscopic data for {\bf 3C226}. Details as in Figure~2.}
\end{figure*}

\begin{figure*}
\centerline{
\psfig{file=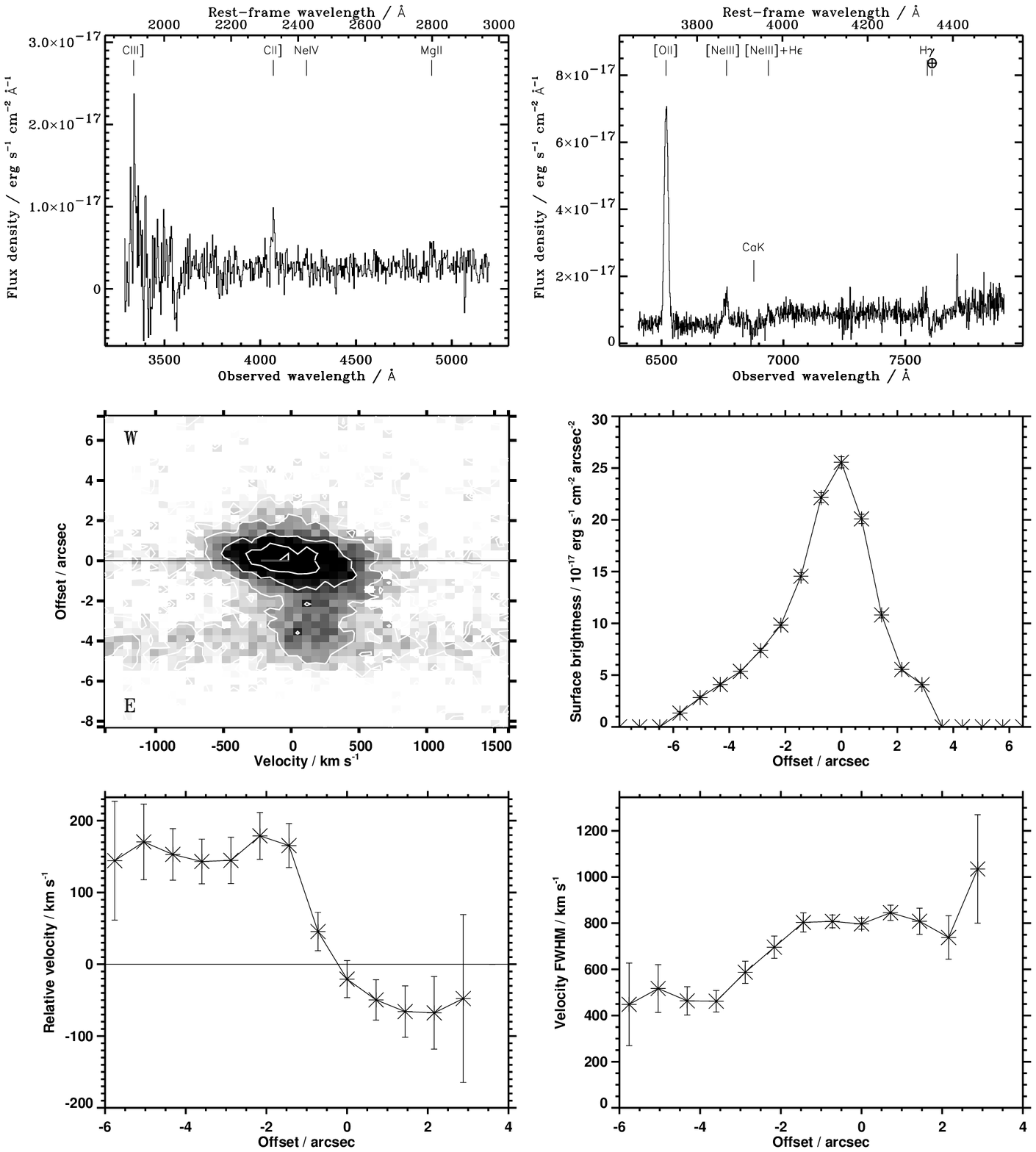,width=\textwidth,clip=}
}
\caption{The spectroscopic data for {\bf 3C247}. Details as in Figure~2.}
\end{figure*}

\begin{figure*}
\centerline{
\psfig{file=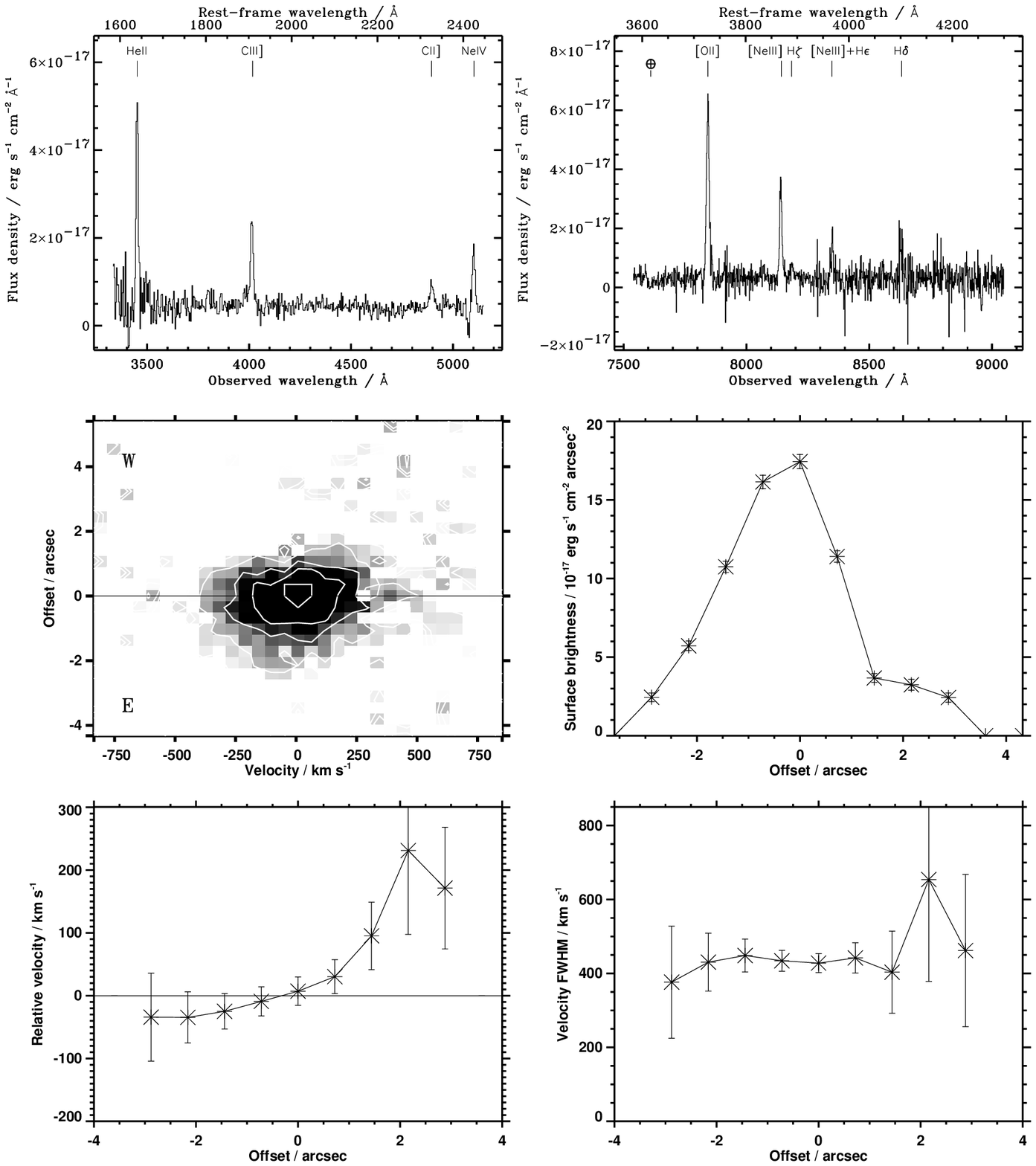,width=\textwidth,clip=}
}
\caption{The spectroscopic data for {\bf 3C252}. Details as in Figure~2.}
\end{figure*}

\begin{figure*}
\centerline{
\psfig{file=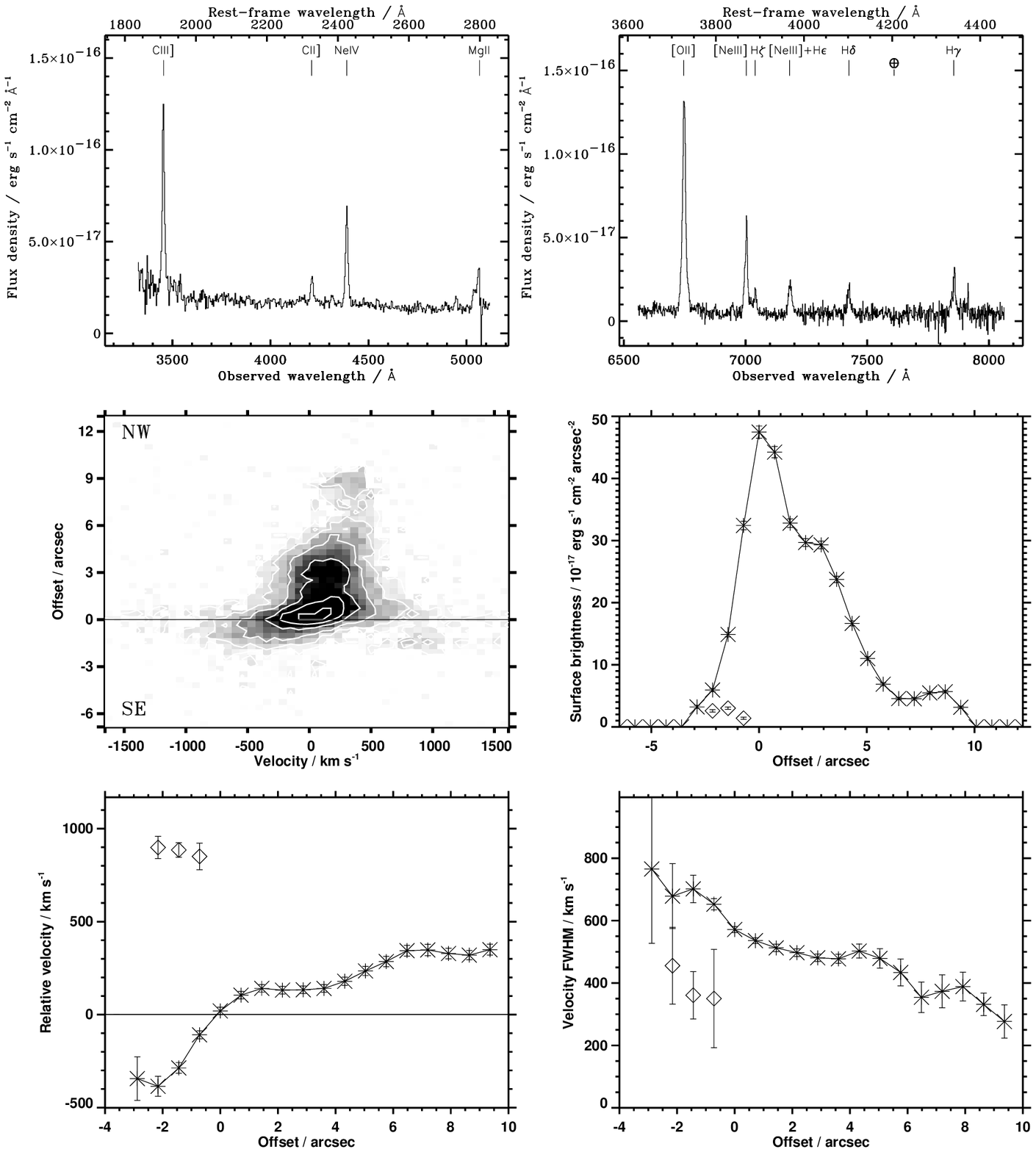,width=\textwidth,clip=}
}
\caption{The spectroscopic data for {\bf 3C265}. Details as in Figure~2.}
\end{figure*}

\begin{figure*}
\centerline{
\psfig{file=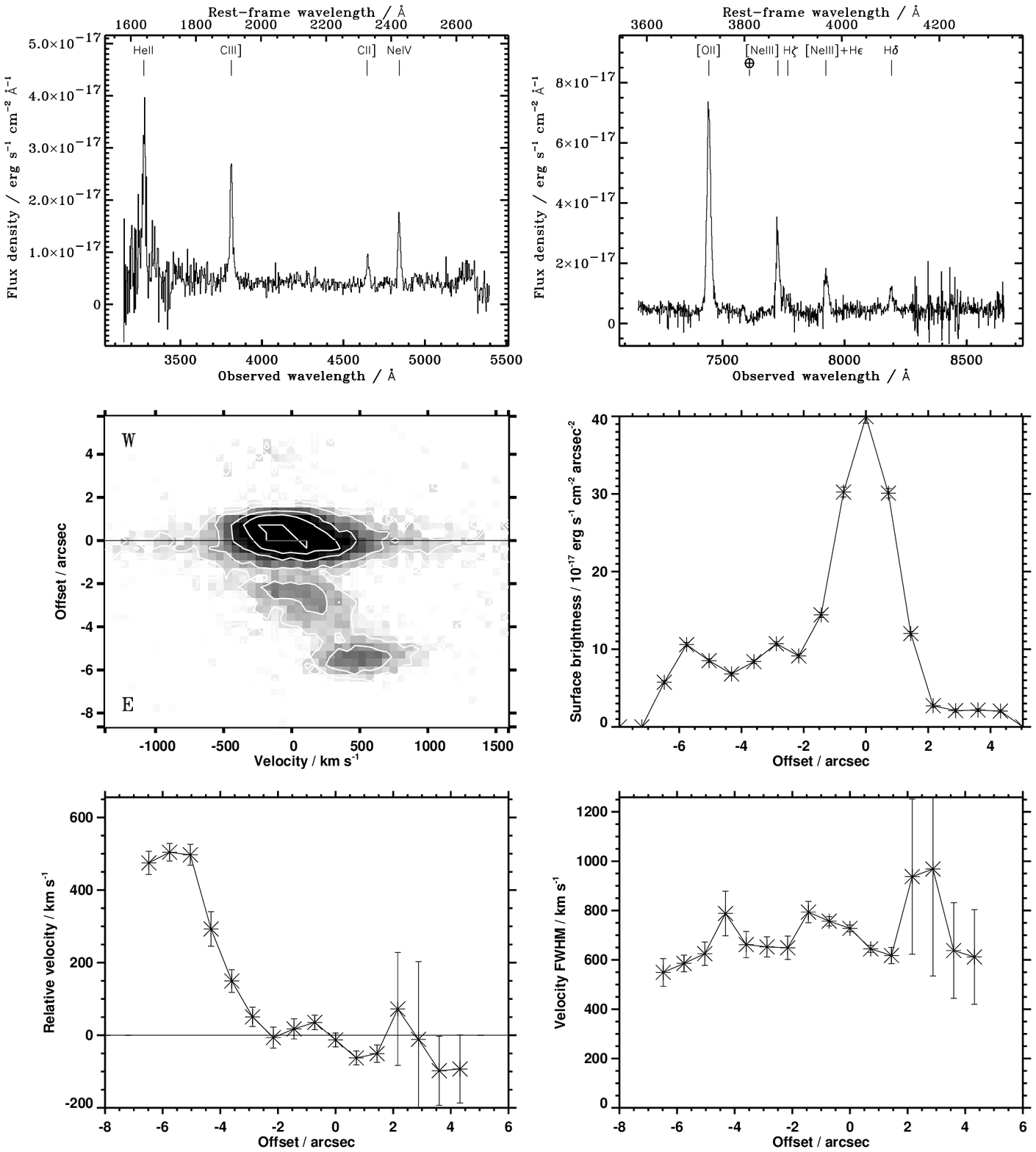,width=\textwidth,clip=}
}
\caption{The spectroscopic data for {\bf 3C280}. Details as in Figure~2.}
\end{figure*}

\begin{figure*}
\centerline{
\psfig{file=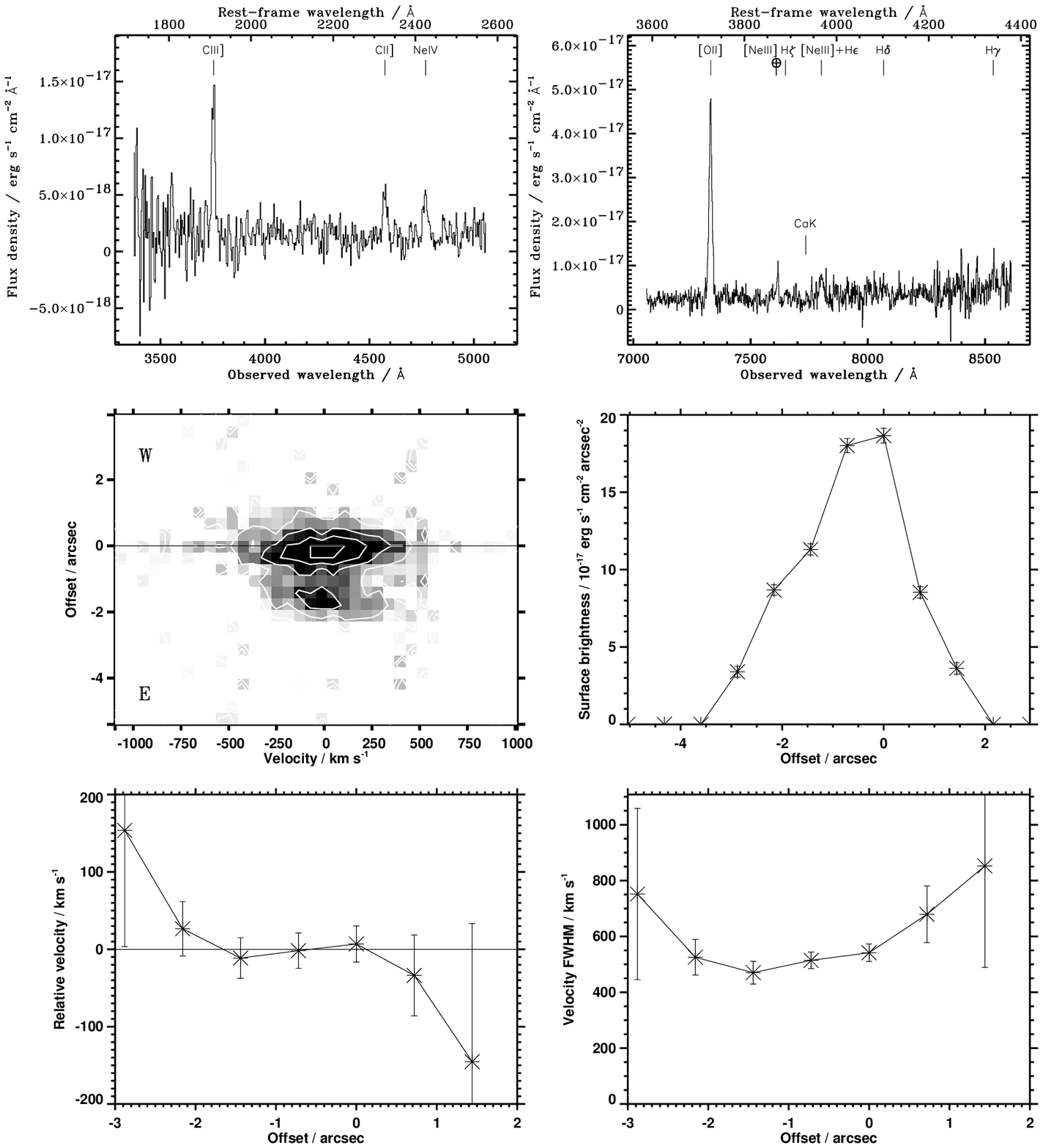,width=\textwidth,clip=}
}
\caption{The spectroscopic data for {\bf 3C289}. Details as in Figure~2.}
\end{figure*}

\begin{figure*}
\centerline{
\psfig{file=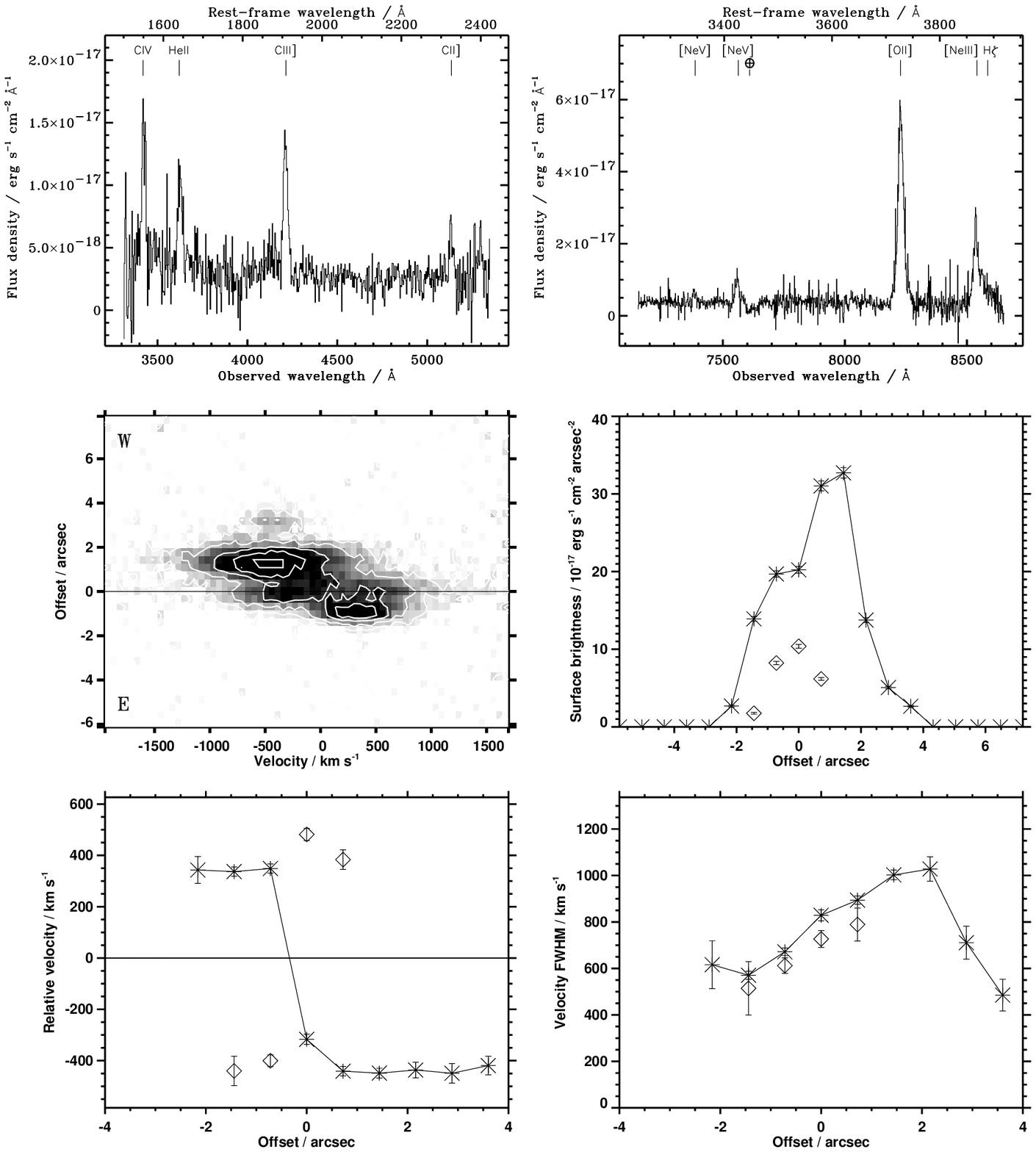,width=\textwidth,clip=}
}
\caption{The spectroscopic data for {\bf 3C324}. Details as in Figure~2.}
\end{figure*}

\begin{figure*}
\centerline{
\psfig{file=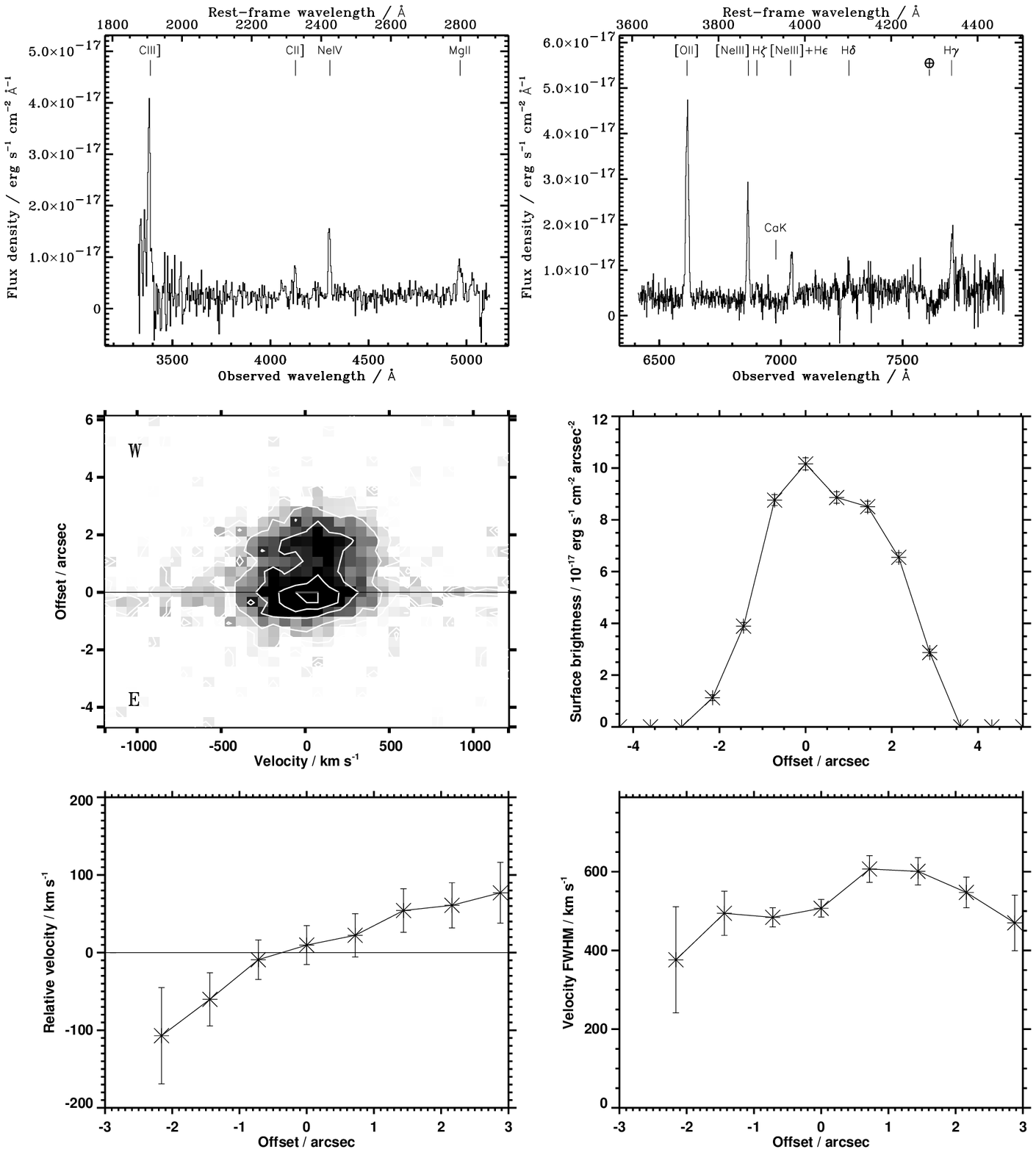,width=\textwidth,clip=}
}
\caption{The spectroscopic data for {\bf 3C340}. Details as in Figure~2.}
\end{figure*}

\begin{figure*}
\centerline{
\psfig{file=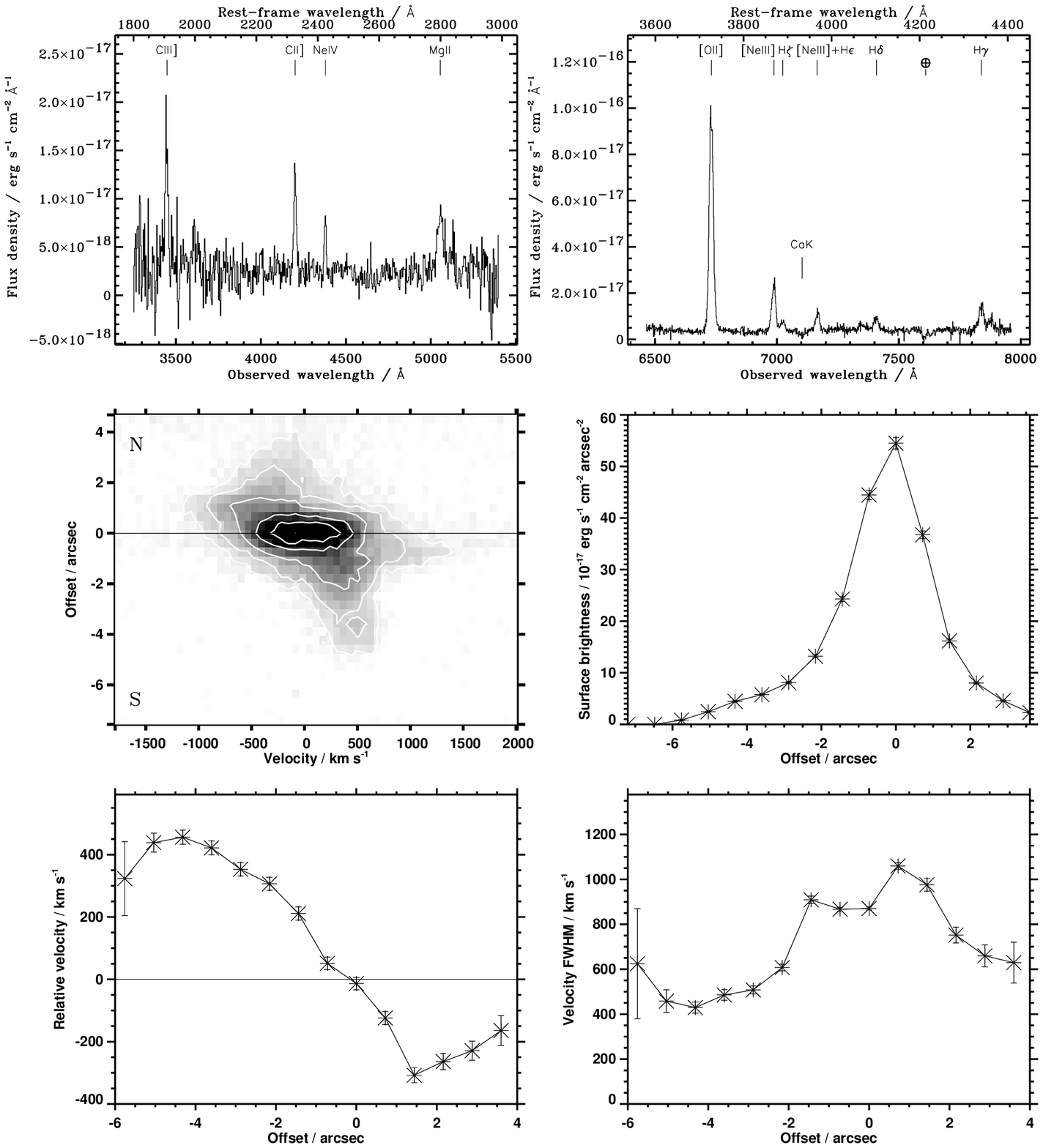,width=\textwidth,clip=}
}
\caption{The spectroscopic data for {\bf 3C352}. Details as in Figure~2.}
\end{figure*}

\begin{figure*}
\centerline{
\psfig{file=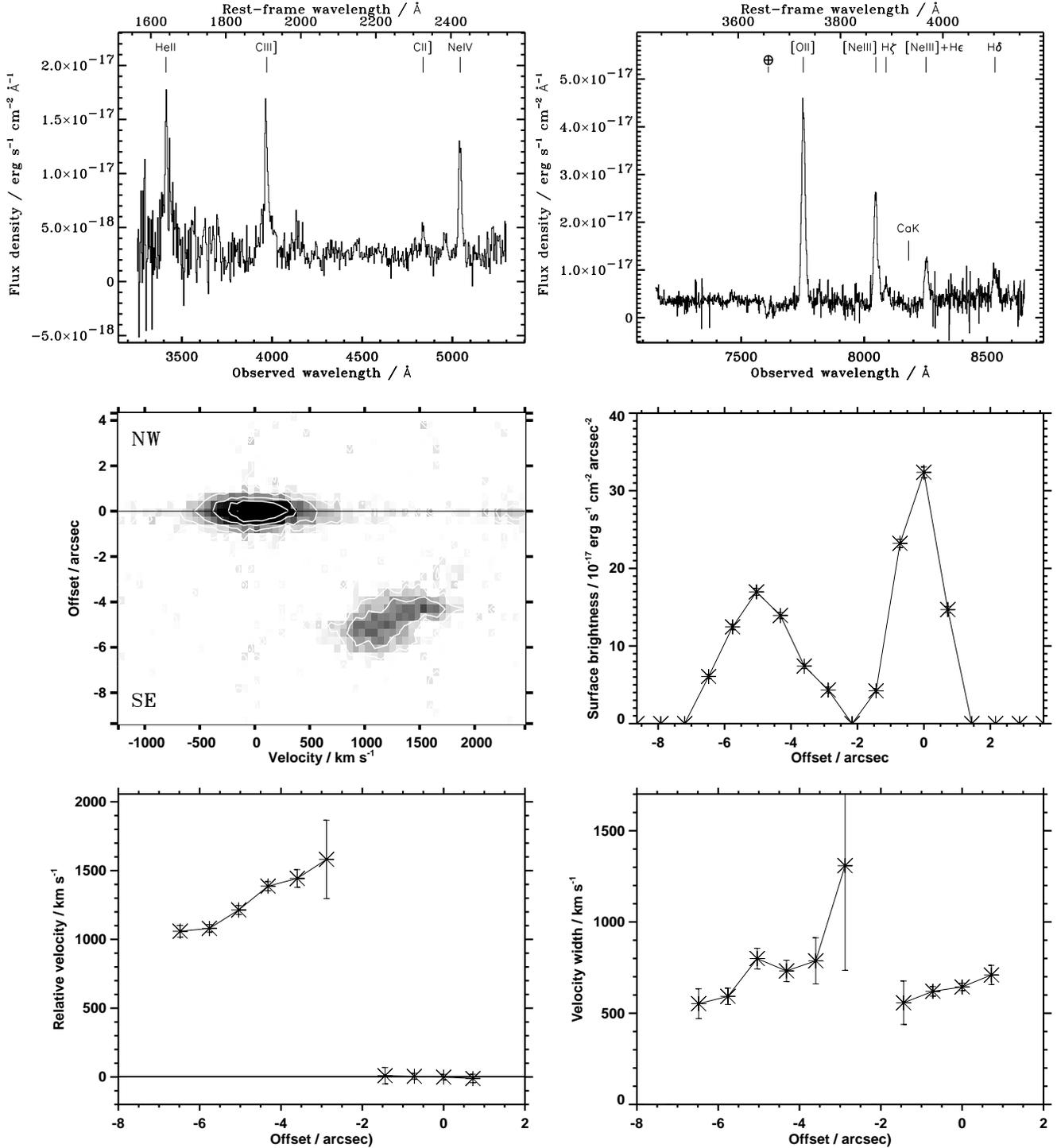,width=\textwidth,clip=}
}
\caption{The spectroscopic data for {\bf 3C356}. Details as in
Figure~2. The plotted one-dimensional spectra are for the northern
galaxy.}
\end{figure*}

\begin{figure*}
\centerline{
\psfig{file=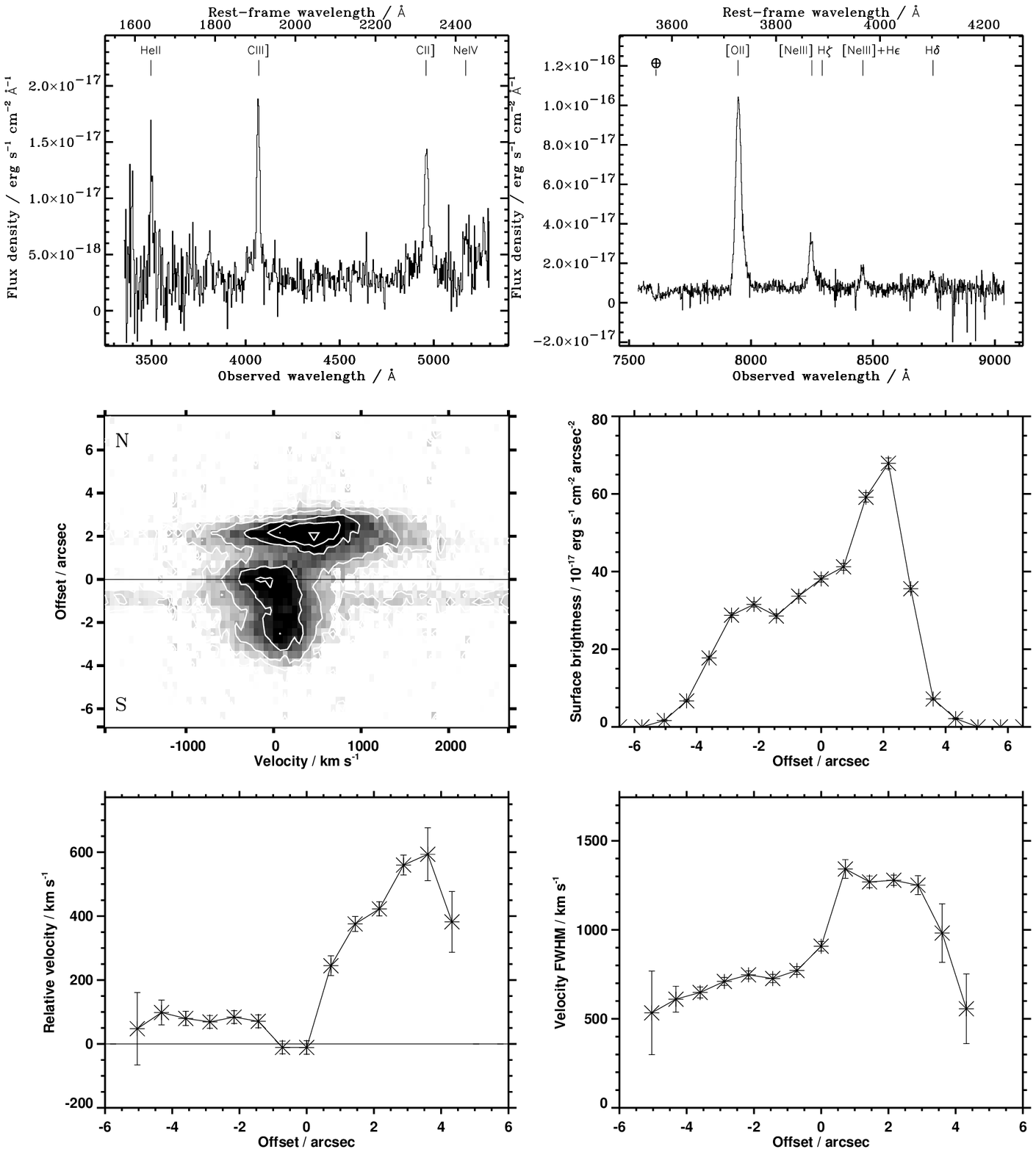,width=\textwidth,clip=}
}
\caption{The spectroscopic data for {\bf 3C368}. Details as in Figure~2.}
\end{figure*}

\begin{figure*}
\centerline{
\psfig{file=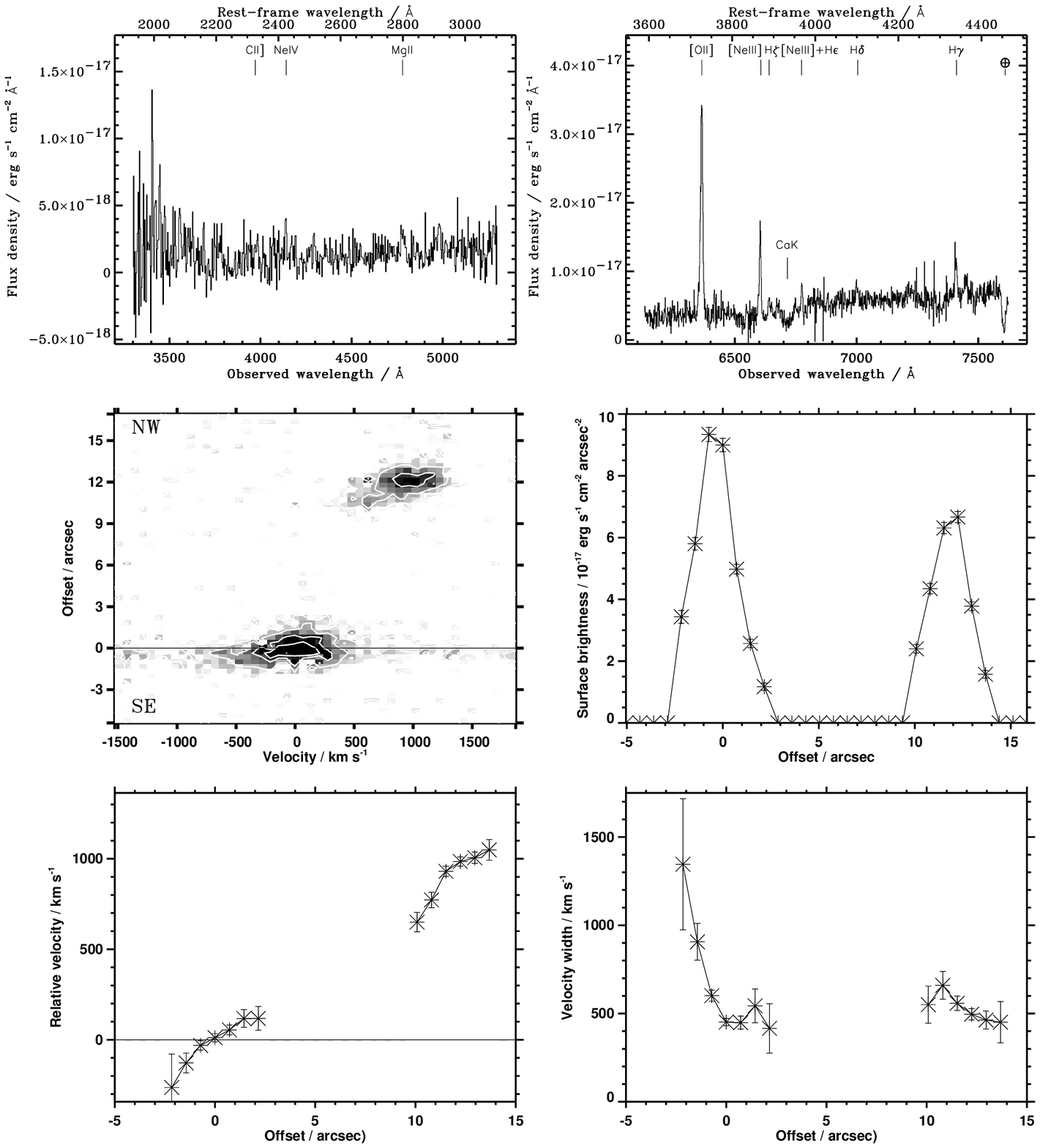,width=\textwidth,clip=}
}
\caption{The spectroscopic data for {\bf 3C441}. Details as in Figure~2.}
\end{figure*}

Also presented in Table~2 are the observed strengths of the 4000\AA\
break, as determined by the ratio of the mean {\it continuum} flux between
4050\AA\ and 4250\AA\ to that between 3750\AA\ and 3950\AA\ \cite{bru83};
note that due to the presence of the excess aligned optical--UV emission,
the strength of this break cannot be used directly to age the stellar
populations of these galaxies, although the galaxies with only weak
alignment effects (e.g. 3C441) do show fairly strong breaks indicative of
evolved stellar populations. 

To study the velocity structure of the [OII]~3727 emission line, a
two--dimensional region around this emission line was extracted
[Figures 2 to 15(c)], and from this a series of one dimensional
spectra were extracted from spatial regions of width 4 pixels (1.44
arcsec), with the extraction centre stepped in units of 2 pixels (0.72
arcsec, $\sim \frac{2}{3}$ of a seeing profile). Each extracted
spectrum was then analysed using the following automated procedure.

\begin{enumerate}
\item The extracted spectrum was fitted to find the best--fitting
Gaussian, allowing for continuum subtraction. If this had a velocity
FWHM greater than the instrumental resolution\footnote{The exclusion
on the basis of FWHM was necessary to avoid selection of single pixel
spikes, but it should be noted that real features may have a measured
FWHM less than the velocity resolution, and thus be excluded, if they
are detected with only low signal--to--noise. However, all of the
extracted Gaussian profiles are found to have deconvolved FWHM in
excess of 200\kms (Figures~2 to 15f) and so it is extremely unlikely
that any real features have been excluded by this method.}, determined
by measuring the FWHM of unblended sky lines, and had an integrated
signal--to--noise ratio greater than 5 then it was accepted. Otherwise
no fit was made at this spatial position.

\item The spectrum was then fitted using a combination of two
Gaussians. This fit was preferred to the single Gaussian fit only if
both fitted Gaussians were wider than the velocity resolution, had an
integrated signal--to--noise in excess of five, and the reduced
chi--squared of two Gaussian fit was below that of the single Gaussian
fit. If these requirements were not satisfied, the single Gaussian fit
was adopted. Note that the amplitude of a fitted Gaussian was allowed
to be negative to detect absorption features (although none were
observed).

\item This process was repeated using 3,4,5 etc Gaussians.
\end{enumerate}

In this way, it was possible to search for high velocity gas components,
and structures in the emission line gas inconsistent with being a single
velocity component (e.g. see 3C324; Figure~10). For each extracted
Gaussian, the integrated emission line flux was calculated, as was the
velocity relative to that at the centre of the galaxy and the FWHM of the
emission line. The last of these was deconvolved by subtracting in
quadrature the instrumental FWHM, as determined from unblended sky
lines. The errors on each of these three parameters were also
determined. It should be noted that a Gaussian did not always provide an
ideal fit to the velocity profile, for example with profiles showing
slight wings in either the blue or red direction, perhaps associated with
a weaker emission component at a different velocity that was too faint to
be individually distinguished.

The variation of the intensity, velocity and FWHM of the [OII]~3727 line
emission with spatial location along the slit are presented in Figures~2
to 15(d to f). The large--scale variations in these three parameters
measured here agree extremely well with those determined from lower
spatial and spectral resolution data by McCarthy \etal\ \shortcite{mcc96a}
for the seven galaxies in common between the two samples; the only
significant exception is 3C324, where the higher spectral resolution of
the current data has shown that a single Gaussian component is clearly not
sufficient to describe the velocity structure. Note also that the surface
brightnesses of the [OII] emission line determined from these spectra are
comparable to those measured from the same region in narrow--band imaging
of this emission line by McCarthy \etal\ \shortcite{mcc95}.

Important features of the emission line properties of individual galaxies
are discussed briefly below. A full discussion of the continuum
morphologies of these sources can be found in Best \etal\
\shortcite{bes97c}, and is not repeated here except where of direct
relevance. 

\medskip
 
{\bf 3C22} has been identified as possessing a significant quasar
component on the basis of a broad \Ha\ emission line and its high
luminosity and nucleated appearance in the K--band
\cite{dun93,raw95,eco95}. The emission line properties observed here,
however, are by no means extreme (Figure~2).  The [OII] line emission is
confined to approximately the inner 2 arcsecond ($\approx 17$\,kpc) radial
distance along the slit (see also McCarthy \etal\ 1995)\nocite{mcc95} and
is consistent with velocity variations $\lta 100$\kms. The FWHM seen for
this line is high (700 to 800\kms) but not exceptional with respect
to the rest of the sample. The ratio of emission line fluxes seen from
this galaxy are intermediate within the sample, and similar to the
combined radio galaxy spectrum of McCarthy \shortcite{mcc93}. The
continuum emission at rest--frame wavelengths $\lta 3000$\AA\ is somewhat
bluer than average.

\medskip

{\bf 3C217} possesses by far the highest equivalent width [OII]~3727 line
emission of all of the galaxies in this sample.  This intense line
emission is relatively compact (Figure~3; see also the narrow band
[OII]~3727 image of Rigler \etal\ 1992)\nocite{rig92} and confined to the
inner 2--3 arcsec radius, in the region in which the HST images also show
very luminous and blue rest--frame ultra--violet emission \cite{bes97c}.
The [OII] line shows a large velocity dispersion and a complex velocity
profile, but with only small ($\lta 200$\kms) variations along the slit in
the mean velocity. Relative to the other galaxies in the sample, the lower
ionisation lines are strong in the spectrum of this object.

\medskip

{\bf 3C226} shows a smooth, regular emission line gas profile
(Figure~4). The [OII]~3727 emission line shows a clear intensity peak at
the centre of the galaxy, extended slightly to the north--west (see also
the narrow--band image of McCarthy \etal\ 1995)\nocite{mcc95} where the
HST image shows a faint blue knot of emission \cite{bes97c}. The relative
velocity plot is consistent with a simple rotating halo; it may instead
represent material infalling or outflowing along the radio axis, although
the smooth slope of the velocity profile would be surprising in that case.
The FWHM of the line profile is low relative to the other sources in the
sample and fairly constant along the slit. At any given location along the
slit, however, the dispersion in the line velocities is far greater than
the mean offset velocity of the emission line at that location, indicating
that whether the relative velocity plot represents a mean rotational
motion or if it arises through inflow or outflow of material, then there
is considerable scatter in the emission line cloud velocities relative to
these mean motions. The emission line ratios of 3C226 are fairly typical
for the sample; the continuum emission in the blue--arm spectrum is redder
than the average.

\medskip

{\bf 3C247} has line emission extending for over 10 arcseconds along the
radio axis (Figure~5; see also McCarthy \etal\ 1995). The inner
approximately 2 arcsec radius of the line emission is almost symmetrical,
with an intermediate velocity FWHM and a velocity profile which again may
be consistent with a mean rotational motion or with infall\,/\,outflow of
material. Further to the north--east there is a smooth transition into a
region of [OII] emission redshifted by 150\kms and with a lower velocity
width. This second region has an associated continuum object, and it seems
likely that what is seen here is an interaction of the radio galaxy with a
companion. The radio galaxy itself shows a significant 4000\AA\ break of
strength $1.61 \pm 0.06$; bearing in mind that the true strength of this
break is diluted by aligned continuum emission, the host galaxy must
contain a well--evolved stellar population. A strong CaK~3933\AA\
absorption feature is readily apparent in the red--arm spectrum
(Figure~5b).

\medskip

{\bf 3C252} shows line emission extended over only a few arcseconds, with
a smooth velocity profile again representing simple rotation or
infall\,/\,outflow (Figure~6). The galaxy has the lowest velocity FWHM of
any source in the sample, although still with a velocity dispersion
significantly greater than the mean relative velocities. The integrated
[OII]~3727 flux is amongst the lowest in the sample, but many of the other
emission lines are strong by comparison.

\medskip

{\bf 3C265} is an extreme radio galaxy in both its continuum and emission
line properties. More than a magnitude brighter at optical wavelengths
than other radio galaxies at the same redshift, its continuum emission is
composed of a large number of components extending over 80\,kpc (10
arcsec) with a remarkably blue colour (Figure~7a); its [OII]~3727
emission shows a similar, or even greater, extent (Figure~7; see also
Tadhunter 1991; Rigler \etal\ 1992; McCarthy \etal\ 1995, 1996; Dey \&
Spinrad 1996)\nocite{tad91,rig92,dey96,mcc95,mcc96a}.  From the galaxy
centre the emission extends a considerable distance to the north--west
with a fairly flat velocity profile and decreasing velocity width. The
`blob' of line emission offset 9 arcsec to the north--west of the galaxy
centre is associated with a continuum emission region (e.g. see Best
\etal\ 1997). The properties of the [OII]~3727 emission line are also
observed with lower signal--to--noise in weaker emission lines.

Tadhunter \shortcite{tad91} reported the presence of high velocity gas
components to the south--east of the nucleus, with velocities of +750 and
+1550\kms\ with respect to the velocity at the continuum centroid,
although these were not obvious in the data of Dey \& Spinrad
\shortcite{dey96} nor of McCarthy \etal\ \shortcite{mcc96a}. The current
data confirm the presence of the +750\kms component, the slightly higher
velocity measured here being due to a small offset between the continuum
centroid position determined here and that of Tadhunter. This component is
also detected clearly in the [NeIII]~3869 line. The +1550\kms\ component
is, however, not detected; this may be due to the difference in slit
position angle of the two observations (136$^{\circ}$ versus
145$^{\circ}$) and/or the use of a narrower slit in the current
observations. The origin of the high velocity component is almost
certainly related to the radio source activity \cite{tad91}.

\medskip

{\bf 3C280} has a complex emission line structure extending over 11 arcsec
(90\,kpc; Figure~8). The emission shows a strong central peak together
with a large extension to the east where it forms a loop around the
eastern radio lobe \cite{rig92,mcc95}. This loop of emission is redshifted
with respect to the velocity at the continuum centroid by about
500\kms. The FWHM of the [OII]~3727 emission is moderately high and almost
constant throughout the entire extent of the emission.

\medskip

{\bf 3C289} shows a central peak of line emission, with a secondary
emission region a couple of arcseconds to the south--east (Figure~9; see
also Rigler \etal\ 1992), corresponding to a faint emission region on the
HST image of Best \etal\ \shortcite{bes97c}. Both the integrated
[OII]~3727 emission line intensity and the FWHM of the emission line are
relatively low for the sample. The velocity profile could represent
rotation or infall\,/\,outflow of material. A weak CaK~3933\AA\ absorption
line may be present in the red--arm spectrum.

\medskip

{\bf 3C324} has previously been described as showing a velocity shear of
700\kms along the radio axis \cite{spi84a,mcc96a}, but the higher spectral
and spatial resolution data presented in Figure~10 clearly indicate that
that is not the case. The emission line gas is composed of two distinct
components, with velocities separated by $\sim 800$\kms. At the position
corresponding to the continuum centroid, the two velocity components
overlap; the adoption of the mean of these two as the true redshift of the
system is necessarily uncertain, and the possibility that the true centre
of 3C324 lies coincident with either of the components determined here to be
at +400 and $-$400\kms\ cannot be excluded. The western emission line
component is slightly more luminous and has the higher FWHM, reaching over
1000\kms; note that the dissociation of the central emission into two
separate components means that a FWHM as high as 1500\kms, determined by
McCarthy \etal\ \shortcite{mcc96a} for the blended pair, is not measured
here. It is unclear whether these two emission line regions represent
different physical systems, perhaps undergoing a merger, or whether
radial acceleration by radio jet shocks is responsible for the bimodality
of the emission line velocities.

This two component structure of the emission line properties of 3C324
reflects the structure of its optical--UV continuum emission
\cite{lon95,dic96}. The HST images show bright emission regions to the
east and west, but a central minimum corresponding to the radio core
position and interpreted as extinction by a central dust lane.
Narrow--band images of the [OII]~3727 emission line also show an elongated
clumpy morphology \cite{ham90,rig92}.  Cimatti \etal\ \shortcite{cim96}
showed that the polarisation properties of the emission to the east and the
west of the nucleus also differ strongly.

\medskip

{\bf 3C340} is another radio galaxy whose emission line structure is
smooth and well--ordered (Figure~11). The relative velocity plot is
consistent with simple rotation or infall or outflow of material, and the
line widths are the second lowest in the sample. The emission is centrally
concentrated, with a small (2 to 3 arcsec) extension along the radio axis
to the west (see also the narrow--band image of McCarthy \etal\ 1995). The
integrated [OII]~3727 intensity is relatively low, with the emission line
ratios in the spectrum indicating a very high ionisation state. The galaxy
shows a significant 4000\AA\ break ($1.52 \pm 0.07$), a broad CaK~3933\AA\
absorption feature, and a red colour for its short wavelength continuum
emission.

\medskip

{\bf 3C352} shows an elongated [OII]~3727 emission region extending for 10
arcseconds, and possibly further since the presence of a bright star to
the north prohibits the detection of any further line emission in that
direction. The velocity profile is smooth throughout the central regions
of the source with a velocity shear exceeding 700\kms, but distorts
somewhat at larger distances (Figure~12). The FWHM is large, reaching
over 1000\kms. These results are consistent with those of Hippelein and
Meisenheimer \shortcite{hip92} from Fabry--Perot imaging. Relative to the
rest of the sample, the lower ionisation lines are strong in the
spectrum. A broad CaK absorption feature can be seen at 3933\AA.

\medskip

{\bf 3C356} has long been a puzzle, with two equally bright infrared
galaxies separated by about 5 arcsec corresponding to the location to two
radio core--like features. The identification of the true nucleus has been
a matter of some debate, with different authors favouring the northern or
the southern galaxy for different reasons (see Best et~al 1997 for a
more complete discussion). For the current data, the slit was placed to include
both components, with zero offset corresponding to the location of the
northern galaxy (see Figure~13). As is observed for the continuum emission
(e.g. Rigler \etal\ 1992, Best \etal\ 1997), the line emission from the
northern region is compact whilst that from the southern region is more
extended but gives a comparable integrated intensity (see also Lacy \&
Rawlings 1994, McCarthy \etal\ 1996)\nocite{lac94,mcc96a}. The northern
region shows virtually no variation in its velocity with position, and a
low velocity width; the southern region, redshifted by about 1200\kms,
shows a steep velocity shear of 500\kms\ in 3 arcsec and a slightly
broader FWHM.

\medskip

{\bf 3C368} is easily the best--studied galaxy in this sample (e.g. Hammer
\etal\ 1991, Meisenheimer \& Hippelein 1992, Rigler \etal\ 1992, Dickson
\etal\ 1995, Longair \etal\ 1995, McCarthy \etal\ 1995, 1996, Stockton
\etal\ 1996, Best \etal\ 1997, 1998a to name only the most recent),
\nocite{ham91,mei92,dic95,sto96a,lon95,bes98a} showing a highly elongated
morphology in both its continuum and line emission extending about 10
arcsec (Figure~14), comparable to the extent of the radio source. The
velocity structure of the line emission determined here is in full
agreement with the previous lower spectral resolution measurements and the
Fabry--Perot imaging of Meisenheimer \& Hippelein \shortcite{mei92}, the
northern knots showing a velocity offset of order 600\kms\ and an extreme
FWHM (up to 1350\kms). Many of the lower ionisation lines in the spectrum
appear strong relative to the other galaxies in the sample. Study of the
continuum emission is hampered by the presence of a galactic M--dwarf star
lying within a couple of arcseconds of the centre of 3C368. \cite{ham91}.

\medskip

{\bf 3C441} shows, in addition to the [OII]~3727 emission from the central
galaxy, a secondary region of emission offset 12 arcseconds to the
north--west and about 800\kms\ redwards in velocity (Figure~15;
cf. McCarthy \etal\ 1996, Lacy \etal\ 1998). This emission region lies
close to the radio hotspot \cite{mcc95} and has been associated with an
interaction between the radio jet and a companion galaxy to 3C441
\cite{lac98}. The [OII]~3727 emission associated with the host galaxy
itself has a low integrated intensity and a smooth velocity gradient of
nearly 400\kms\ in 5 arcsec, consistent with rotation or with
infalling\,/\,outflowing gas. A relatively large 4000\AA\ break is
observed in the spectrum ($1.64 \pm 0.04$), together with strong
CaK~3933\AA\ absorption, consistent with the fact that this galaxy also
shows only a very weak alignment effect at optical--UV wavelenths
\cite{bes97c}.
\bigskip

\begin{figure}
\centerline{
\psfig{file=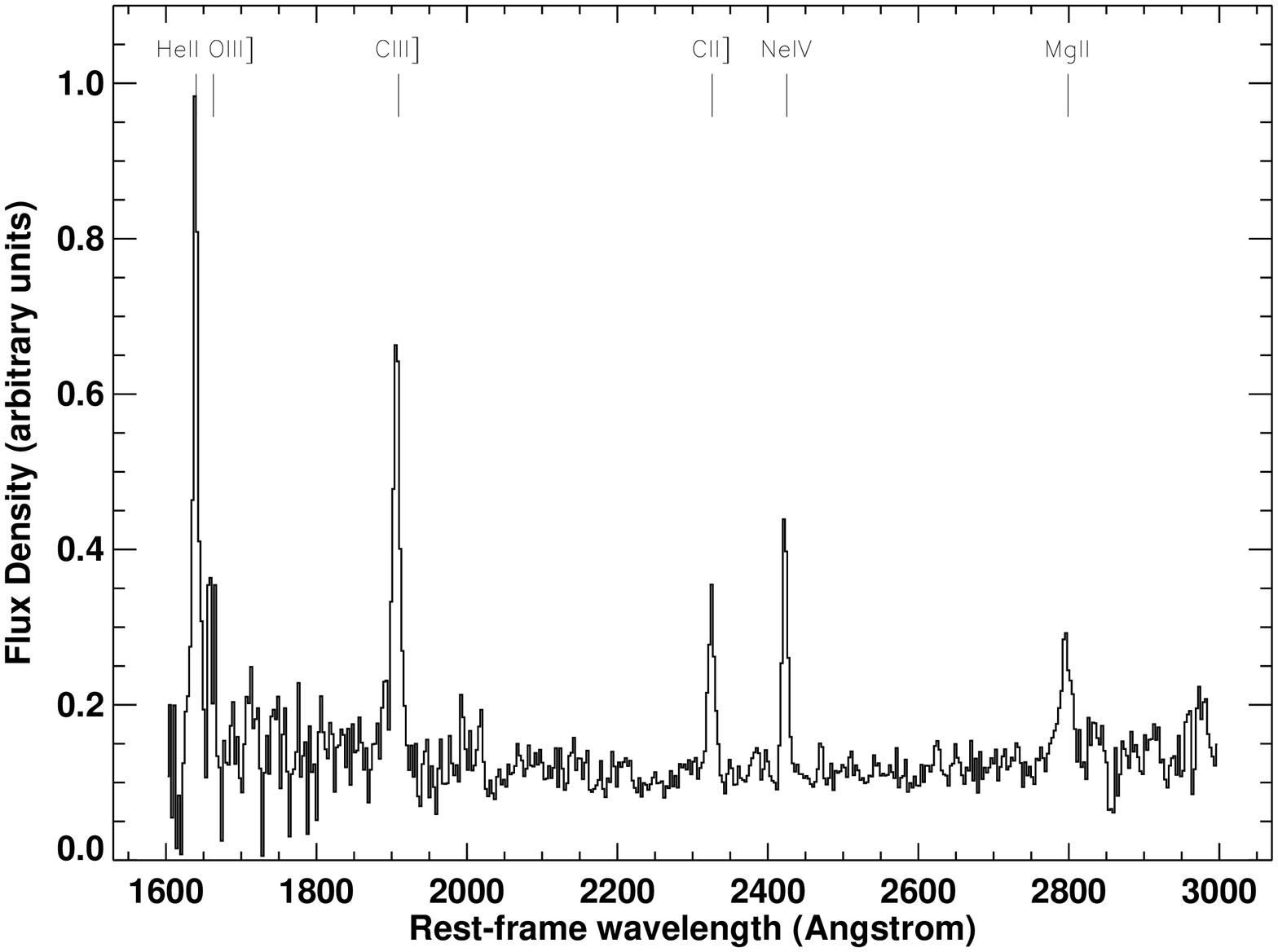,width=7.8cm,angle=90,clip=}
}
\centerline{
\psfig{file=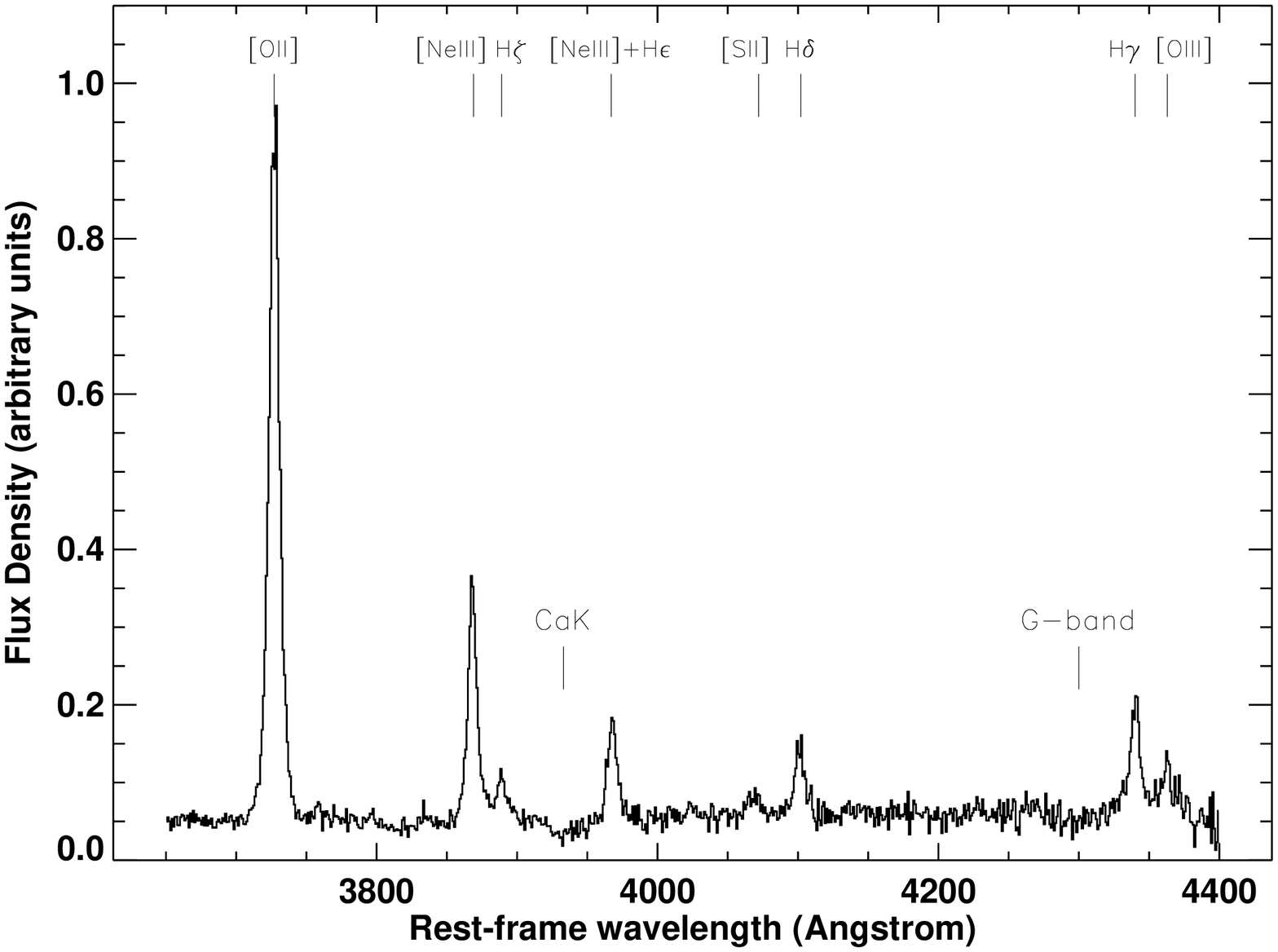,width=7.8cm,angle=90,clip=}
}
\caption{\label{totals} A summation of the spectra for all of the radio
galaxies.}
\end{figure}

Composite spectra have been produced for each of the red and blue arms of
the spectrograph, by combining all of the presented spectra at the same
rest--frame wavelengths, giving each individual spectrum an equal
weighting. 3C368 was excluded from this combined spectrum due to the
contribution of the foreground M--star to its emission. The resulting
total spectra, shown in Figure~\ref{totals}, are equivalent to single
spectra of over 20 hours in duration. In Table~\ref{avlines} are tabulated
the relative strengths of the emission lines in this composite
spectrum. These are quoted relative to the commonly adopted scale of
\Hb$=100$ by assuming \Hc/\Hb$\approx 0.47$, appropriate for Case B
recombination at T=10000\,K \cite{ost89}; this value is also consistent
with that obtained from the \Hd\ line assuming \Hd/\Hb$\approx 0.26$.

One feature is immediately apparent when comparing these relative line
fluxes with those from the composite spectrum of radio galaxies with
redshifts $0.1 < z < 3$ constructed by McCarthy \shortcite{mcc93}: the
emission lines at short wavelengths are less luminous by factors of 2 to 4,
relative to \Hb, than those of McCarthy's spectrum. This may be due to the
wide range of redshifts of the radio galaxies making up McCarthy's
composite and the strong correlation between emission line flux and
redshift \cite{raw91b}; the shortest wavelength lines in his composite
spectrum are only observed in the highest redshift sources (with powerful
line emission) whilst the \Hb\ line is seen in the lower redshift sources,
introducing a bias towards lines at shorter rest--frame wavelengths
appearing more luminous. The composite spectra presented in
Figure~\ref{totals} and Table~\ref{avlines} are much less prone to this
bias, and so provide a fairly accurate measure of relative line fluxes at
redshift $z \sim 1$.

Besides the emission lines, other features visible in the spectra include
the broad CaK absorption feature at 3933\AA, with an equivalent width of
$10 \pm 2$\AA, and a weaker G--band absorption at 4300\AA\ with an
equivalent width of $7 \pm 3$\AA. A 4000\AA\ break is marginally visible,
but there is little evidence for the spectral breaks at 2640 and 2900\AA\
(cf. Spinrad \etal\ 1997)\nocite{spi97} expected from an old stellar
population. This is not too surprising since the contribution from the old
stars to the total flux density at these wavelengths, and indeed
throughout all of the combined blue arm spectrum, is small compared to
that of the aligned emission.

\begin{table}
\caption{\label{avlines} Emission line fluxes of the `average' spectrum
of the 3CR radio galaxies, relative to H$\beta = 100$.}
\begin{center}
\begin{tabular}{lcclc}
~~~~Line     &Flux&     & ~~~~Line                    &Flux\\
HeII 1640    & 95 &~~~~~& [NeIII] 3869                & 65 \\
OIII] 1663   & 19 &     & H$\zeta$ 3889               & 12 \\
CIII] 1909   & 83 &     & H$\epsilon$ + $[$NeIII] 3967& 26 \\
CII] 2326    & 27 &     & [SII] 4072                  & 9  \\  
$[$NeIV] 2425& 33 &     & H$\delta$ 4102              & 24 \\
MgII 2798    & 38 &     & H$\gamma$ 4340              & 47 \\
$[$OII] 3727 & 256&     & [OIII] 4363                 & 21 \\
\end{tabular}
\end{center}
\end{table}

\section{Conclusions}
\label{conc}

Extremely deep spectroscopic observations have been presented of an
unbiased sample of the most powerful radio galaxies with redshifts $z \sim
1$. A broad range of emission lines is seen and a study at intermediate
spectral resolution of the two--dimensional velocity structures of the
emission line gas are presented. The enhanced sensitivity of new CCDs at
short wavelengths has enabled the measurement of emission line ratios and
continuum flux densities at unprecedentedly short wavelengths, $\lambda
\lta 3500$\AA, corresponding to the near--UV in the rest--frame of the
sources where any continuum contribution from an evolved stellar
population will be negligible.

The main results can be summarised as follows:

\begin{itemize}
\item Analysis of the velocity structures of these galaxies shows them to
exhibit a wide range of kinematics. Some sources have highly distorted
velocity profiles and velocity FWHM exceeding 1000\kms. Other sources have
lower velocity dispersions and more ordered emission line profiles, with
the variation of mean velocity along the slit being consistent with simple
rotation. Even in these latter sources, however, the velocity FWHM are
still a few hundred \kms, significantly larger than the variations in mean
velocities, indicating that there is considerable scatter in the emission
line cloud velocities relative to any mean rotational motion.

\item A high velocity ($\sim 750$\kms) gas component is confirmed close to
the nucleus of 3C265. This is unique amongst the sample, but other
galaxies display gas with velocities $\ga 400$\kms\ offset a few arcseconds
from the centre, either connected to the central emission line region
(3C280, 3C352, 3C368) or as a discrete region (3C356, 3C441). 

\item 3C324 is shown to consist of two kinematically distinct components
separated in velocity by 800\kms. 

\item For those galaxies in which the alignment effect is seen to be
relatively weak in the HST images, and hence the spectra are not dominated
by emission from these alignment processes, 4000\AA\ breaks from evolved
stellar populations are clearly visible. CaK absorption features are also
readily apparent in a number of the spectra.

\item At rest--frame wavelengths shortward of $\sim 2500$\AA\ the
continuum emission of the galaxies is, on average, relatively flat in
$f_{\lambda}$, although considerable source to source variations are seen
both in these continuum colours and in the emission line ratios.

\item A composite spectrum gives the relative strengths of the emission
lines at rest--frame wavelengths between HeII~1640 and [OIII]~4363.
Emission lines at short rest--frame wavelengths are systematically weaker
(relative to \Hb) than those in the composite spectrum of McCarthy
\shortcite{mcc93}. It is suspected that this is due to a bias introduced
in McCarthy's spectrum by the emission line strength versus redshift
correlation, and the large redshift coverage of the radio galaxies which
comprise his sample.
\end{itemize}

The broad variation in kinematical and ionisation properties within the
sample as a whole are investigated and compared against other radio source
properties in the accompanying Paper 2, and conclusions are drawn there 
concerning the origin of the ionisation and kinematics of the emission
line gas.

\section*{Acknowledgements} 

This work was supported in part by the Formation and Evolution of Galaxies
network set up by the European Commission under contract ERB FMRX--
CT96--086 of its TMR programme. The William Herschel Telescope is operated
on the island of La Palma by the Isaac Newton Group in the Spanish
Observatorio del Roches de los Muchachos of the Instituto de Astrofisica
de Canarias. We thank the referee, Mike Dopita, for his careful
consideration of the original manuscript and a number of useful
suggestions. 

\bibliography{pnb} 
\bibliographystyle{mn} 
\label{lastpage}

\end{document}